\documentclass{aastex}
\usepackage{spr-astr-addons}
\usepackage{url}
\usepackage{graphicx}
\usepackage{natbib}
\usepackage{hyperref}
\RequirePackage{color}

\maketitle

\begin{document}

\title{Interacting planetary nebulae III: Verification and galactic population based on the measurements of Gaia EDR3.}
\slugcomment{Not to appear in Nonlearned J., 45.}
\shorttitle{Interacting planetary nebulae III}
\shortauthors{Mohery et al.}

\author{Mohery, M.\altaffilmark{1}} \and
\affil{Dept. of Phys, College of Science, University of Jeddah, Jeddah, Saudi Arabia.}
\email{mmohery@uj.edu.sa}
\author{Ali, A.\altaffilmark{2}}
\affil{Astronomy, Space Science \& Meteorology Department, Faculty of Science, Cairo University, Giza 12613, Egypt.}
\email{afouad@sci.cu.edu.eg}
\author{Mindil, A.\altaffilmark{1}}
\and
\author{Alghamdi, S.A.\altaffilmark{3}}
\affil{Astronomy and Space Science Dept., Faculty of Science, King Abdulaziz University, 21589 Jeddah, Saudi Arabia}


\begin{abstract}
The phenomenon of interaction between planetary nebulae (PNe) and the interstellar medium (ISM) is one of the significant issues in the field of astrophysics. The main objective of this paper is to verify the interaction process for objects that have been known as interacting PNe (IPNe) in the literature. This study is based on parallax and proper motion observations facilitated recently by the early third data release of the Gaia space mission. Based on the proper nebular central star (CS) motion towards the region of interaction between the PN and ISM, we were able to verify the interaction process for a group of 68 PNe and disprove the interaction process for a group of 33 PNe. The members of both groups were confirmed as genuine PN-ISM interacting objects in the literature. The members belonging to the 33 PNe group are false PN-ISM interacting objects that mimic the structure of IPNe. Moreover, we calculated the physical and kinematic properties of the verified group and analyzed their galactic population classification using reliable and precise measurements of the proper motion and parallax of the CS. We find that 41\% and 41\% of this group are associated with galactic thin and thick disks, respectively, while 18\% are members of thin or thick disks. The kinematical results show that the galactic thin-disk members have smaller vertical galactic heights, space velocities, and peculiar velocities than those belonging to the galactic thick disk.
\end{abstract}

\keywords{ISM; Planetary nebulae; parallaxes; proper motion}


\section{Introduction}\label{section1}
The mass loss of low and intermediate stars during the planetary nebula stage has a significant impact on the galactic chemical composition and evolution, as it enriches the interstellar medium with nucleosynthesis elements formed in the interiors of the nebular precursor stars. During the evolution of the PN, an interaction with the ISM may occur, resulting in the formation of a bow shock that prevents the nebular shell from expanding in the direction of interaction, increasing the brightness of this region. This depends on the relative velocity and density between the PN and the ambient ISM. {\it Therefore, the direct sign of the PN-ISM interaction is the central star motion towards the interacting region.}

\citet{Ali12}, hereafter Paper I, classified the IPNe into two categories based on the evidence for interaction: confirmed and possible. Objects with more than one sign of interaction (such as asymmetry of the outer nebular region; flux improvement and decrease in ionization level of the interaction region; filaments caused by shock compression due to the ISM) are regarded as confirmed IPNe, but those that are presented as IPNe by \citet{Ali2000} due to their one-sided appearance and/or classified as low-confidence IPNe by \citet{Soker97} and have no other symptom of interaction are considered as possible IPNe. The physical characteristics, galactic distribution, galactic orientation, and morphological classification of the confirmed IPNe group (117 PNe) are discussed in Paper I. They reveal that the majority of IPNe are close to the galactic plane at a vertical height of $\leq 100$ pc, suggesting that they could interact with the molecular and cold neutral clouds. Moreover, they found that interacting regions tend to be parallel to the galactic plane.

\citet{Ali13}, hereafter Paper II, discussed the classification of IPNe in terms of galactic population using a sample of 34 PNe with available proper motions, radial velocities, and distances in the literature. As summarized in Paper I and Paper II, the interaction of PNe with the ISM is a useful phenomenon for understanding many ISM and PN characteristics, including (1) exploring the ISM structure on a small scale; (2) detecting some of the ISM physical parameters, such as density and magnetic field; (3) predicting the space direction of the PN; (4) interpreting the evolution of PNe, particularly in their late stages; (5) explaining the uncommon features in the nebular halos; and (6) understanding the "missing mass phenomenon", which is a key tool for investigating the chemical evolution of a galaxy.

The interaction could happen at any time during the nebular evolution. According to the 3D hydrodynamic simulations presented by \citet{Wareing07}, the PN-ISM interaction can occur early, when the AGB wind (PN extended halo) interacts with the ISM (first stage - WZO\,1), or later, when the ionized shell interacts with the ISM (second stage - WZO\,2). In the late stage, the expansion of the PN shell stops, and its central star moves off its center (third stage - WZO\,3), whereas in its very late stage, the PN appears fully disrupted, with its core star located outside the PN (fourth stage - WZO\,4). In this stage, the material is removed by ram pressure at the PN–ISM interface, forming a tail behind the PN.

The observations of PN-ISM interaction are not restricted to the optical band but have also been extended to the infrared, particularly for the early interaction between the optically faint AGB halo and the ISM (e.g., \citet{Ramos18} and \citet{Hsia20}). Based on the well-defined arc-shaped filament in the southern halo of M\,2–55, identified in H$\alpha$ and mid-infrared observations, \citet{Hsia20} reported that the entire nebula is moving roughly toward the south and probably interacts with its surrounding ISM. Moreover, the highly asymmetric halo seen around NGC\,6772 in the H2 emission line indicates that the object is interacting with the ISM \citep{Fang18}. \citet{Kameswara18} confirm \citet{Martin02}'s prior suggestion of the interaction between NGC\,40 and ISM, where they detected a far ultraviolet (FUV) halo on the trailing side of the nebular head that is oriented in the proper motion direction. \citet{Ramos18} discussed the interaction of the halo surrounding the butterfly PN NGC\,650 with the ISM using high-resolution broad- and narrow-band optical images. They concluded, based on the nebular proper motion and distance, that both the main nebula and the halo travel at a velocity of $\sim 34$ km\,s$^{-1}$ in the SE direction, squeezing the eastern part of the halo into a bow-shock structure. Through their radio investigation of the region surrounding the PN Sh\,2-174, \citet{Ransom15} proposed that the ISM magnetic field is deflected at the nebular leading edge and that the downstream tail and upstream field deflection indicate a PN-ISM interaction. Based on an uncertain nebular distance, the galactic position and velocity components, and thus the galactic population, of this object were computed in both \citet{Ransom15} and Paper II. Thanks to Gaia's precision measurements of parallax and proper motion, we will be able to improve the galactic population of this object and others. \citet{Ali15b} obtained integral field spectroscopic investigation of unstudied evolved PN G342.0-01.7, which shows evidence of interaction with its surrounding ISM. They prove that the nebula is moving in the southeast direction. Furthermore, and based on constructing a shock model, they suggest that the interaction accounts for only $\sim 10\%$ of the total luminosity but has a significant impact on the global spectrum of the PN.

Gaia is a European Space Agency (ESA) space mission that was launched and that is now operating to generate a detailed three-dimensional map of the Milky Way galaxy. The early third data release (Gaia EDR3) appeared in December 2020. The full release of Gaia data (Gaia DR3) is anticipated for the first half of 2022. Gaia EDR3 calculates the position and apparent magnitude of $\sim 1.8$ billion sources and the parallax ($\pi$) and proper motion ($\mu$) of $\sim 1.5$ billion sources. Compared to Gaia DR2, the precision of the stellar parallax and proper motion in this release were enhanced by $30\%$ and $200\%$, respectively \citep{Gaia_Collaboration20}. The estimated parallax zero-point (-0.017\,mas) for Gaia EDR3 data, \citep{Lindegren21}, was enhanced compared to that of Gaia DR2.

The aims of this work can be summarized as follows: (1) verification of the interaction between PNe and ISM based on the direct sign of the direction of interaction should be compatible with the proper nebular central star motion; (2) improvement of the physical and kinematical characteristics of verified IPNe according to the precise data of Gaia EDR3; (3) determination of the space velocity components of verified IPNe to achieve their galactic population; and (4) determination of the objects that mimic IPNe.

Sections 2 presents the data sample. Section 3 discusses the verification IPNe based on the CSs proper motion. The kinematical characteristics and galactic population of the verified objects are analyzed in Section 4, while the mimic IPNe is discussed in Section 5. The conclusion is given in the last section.

\section{The data sample} \label{The calibration sample}

We searched the Gaia database for the parallax and proper motion of the PN central stars that were classified as confirmed IPNe in Paper I and for other PNe that were detected after that time and regarded as IPNe in the literature, e.g., M\,2–55 \citep{Hsia20}. We detected the parallax and proper motion for a sample of 101 PNe  in the Gaia database.

The extracted proper motion and equatorial coordinate were applied to calculate each CS's position angle and, hence, its associated PN travel direction through the ISM.  On the plane of the sky, the zero position angle refers to the north direction, while the 90$^{\rm o}$ angle refers to the east direction. Based on the results of the position angle of the nebular CSs, we divided the data sample into two groups: verified (68 PNe) and mimic (\textbf{33} PNe). The fundamental parameters of both groups are listed in Table \ref{Table1} and Table \ref{Table5}. In Table \ref{Table1}, the PN name, as given in the SIMBAD database, Gaia EDR3 designation, galactic coordinate, equatorial coordinates, CS parallax, proper motion, G magnitude, and BP-RP color index are given in columns 2, 3, 4-5, 6-7, 8, 9-10, 11 and 12, respectively. The PN angular radius, radial velocity, and expansion velocity are given in columns 13, 14, and 15, respectively. The PN images used in constructing figures 1–5 were extracted from the "Aladin Sky Atlas" developed at CDS, Strasbourg Observatory, France. The angular radius of the true IPNe sample, as well as the distances of IPNe with negative parallax and those with higher parallax uncertainty, were collected from \citet{Frew16}. Moreover the radial velocity and expansion velocity of the entire sample were extracted from \citet{Durand98} and \citet{Acker92}, respectively. For the PNe with unknown expansion velocity, we assumed a mean value of 20 km/s \citep{Weinberger89}.

\section{Verification of the PN-ISM interaction}

There are many symptoms of PN-ISM interaction, such as flux enhancement in the nebular region of interaction, asymmetry of the outer region of the nebula, a decrease in the ionization level of the interaction region, a shift of the CS from the geometric center of the nebula (see Figure \ref{Figure1}), and the appearance of filaments due to shock compression by ISM. Furthermore, \citet{Dgani98} proposed that the Rayleigh-Taylor instability can play a substantial impact not only in shaping the outer halo of the nebula but also in shaping its inner regions because it permits the ISM to flow into the nebular inner parts by breaking up the halo.

The CS motion towards the interaction region (nebular bright rim) provides direct evidence of the interaction process. Prior to the Gaia era, the available reliable measurements for CS proper motion in the literature were confined to a small number of objects with low precision compared to Gaia, such as those reported by \citet{Cudworth74}, \citet{Harris07}, and \citet{Kerber08}. We calculate the position angles (PAs) of CSs proper motion with their uncertainties for the data sample.

Due to the low quality of the PN images extracted from the "Aladin Sky Atlas", the region of PN-ISM interaction is not well defined in many cases. As a result, we considered the verified nebula to have a proper motion axis (which shows the direction of motion of the PN through the ISM) that lies within $\pm 15$ degrees of the imaginary interaction axis (which extends from the CS to the midpoint of the interaction region). In figures 2, 3, 4, 5, and 7, the white star represents the position of the PN central star, while the black arrow refers to the direction of PN movement through the ISM.

It should be noted here that the proper motion of the CS has been corrected for the influence of the ISM motion, caused by Galactic rotation, before calculating the position angle (see \citet{Comeron07}). Table 2 displays the verified objects and their position angles that indicate their motion towards the nebular enhancement regions. In columns 3 and 4 of Table 2, we listed the direction of PN-ISM interaction regions and their references, respectively.  Furthermore, the table shows the size and dynamical age of each PN. The table is divided into four parts according to \citet{Wareing07} classification. The median values of the nebular size and dynamical age of each class agree, in general, with the results presented in Paper I (Table 2) within the error range. Moreover, the median size and dynamical age of the entire sample ($0.43\pm0.09$ pc \&	$2.1\rm{x}10^4\,\pm\,4.3\rm{x}10^3$ yr) are also comparable to the results reported in Paper I.

Unfortunately, we were unable to extract suitable images  from the "Aladin Sky Atlas" for the entire PNe sample. This is due to many of PN-ISM interaction features require long exposure times to appear clearly, such as those given by \citet{Corradi03}, and other features appearing only in the infrared domain, such as those reported by \citet{Ransom15} and \citet{Fang18}. As a result, Figures 2, 3, 4, 5 and 7 are restricted to a limited number of true and mimic IPNe.
Figures 2-5 illustrate the direction of CS motion relative to the ISM (black arrow) of the verified IPNe, where the nebular image sample was divided into four groups according to the classification of \citet{Wareing07}. Figures 2, 3, 4, and 5 show examples of IPNe of WZO\,1, WZO\,2, WZO\,3, and WZO\,3/4 classes, respectively.

In light of explaining the morphology of the PFP 1 nebula that displays a brightness enhancement to the NW rim, \citet{Pierce04} proposed
proposed two possible mechanisms for this interaction. The first is that the PN migrates NW relative to its surrounding ISM and is accompanied by a movement of the CS away from the nebular center, whereas the second is that the nebula expands into a dense ISM region to the NW. Due to the CS candidate's lack of accurate proper motions,\citet{Pierce04} are unable to decide which of these mechanisms is correct. Thanks to Gaia, we detect the real PN blue CS and its proper motion, which shows that the entire nebula is moving  to the south direction (see Figure 7(b)). As a result, we conclude that the second scenario, rather than the first, is responsible for the interaction.

\section{Kinematical characteristics of the verified IPNe}

We calculate the space velocity components $U$, $V$, and $W$ using absolute proper motions, radial velocities, and parallax, where $U$ and $V$ denote the direction toward the galactic center and galactic rotation, respectively,  in the galactic disk, and $W$ denotes the direction perpendicular to the galactic disk. We follow the method provided by \citet{Johnson&Soderblom87} for determining the velocity components and uncertainties. The velocity components are corrected for the local standard of rest (LSR) and galactic standard of rest (GSR). The LSR correction accounts for the Sun's peculiar motion ($U_{\odot}$, $V_{\odot}$, $W_{\odot} = 10.0, 5.25, 7.17\,\rm{km\,s}^{-1}$) \citet{Dehnen98}, whereas the GSR correction accounts for the IAU galactic centric distance of the Sun ($R_{\odot} = 7.6$ kpc) and rotational solar velocity V($R_{\odot}$) = 185\,km\,s$^{-1}$.

In Figure 6, we plot the Toomre diagram \citep{Bensby03} to unveil the galactic population of the verified IPNe. The three semicircular lines indicate the constant space velocity ($V_S = \sqrt {U^2_{LSR}+ V^2_{LSR}+ W^2_{LSR}}$) in steps of 50, 70, and 100 km\,s$^{-1}$. The x-axis indicates the velocity component ($V_{LSR}$) in the galactic rotation direction, whereas the y-axis indicates the velocity component in the galactic center direction ($U_{LSR}$) and the direction perpendicular to the galactic plane ($W_{LSR}$). According to this diagram, the objects of space velocity $V_S \leq 50\,\rm{km\,s^{-1}}$, $200 \geq V_S \geq 70\, \rm{km\,s^{-1}}$, and $V_S \geq 200\,\rm{km\,s^{-1}}$ are members of the galactic thin disk, galactic thick disk, and galactic halo, respectively. The objects of $70 \geq V_S \geq 50\,\rm{km\,s^{-1}}$ are probable thin or thick-disk members. From Figure 6, we find that 19 ($41\%$), 19 ($41\%$), and 8 ($18\%$) objects are located in galactic thin disks, thick disks, and thin/thick disks, respectively. In Paper II, we discussed the galactic population of a sample composed of 34 PNe, where 38\%, 29\% and 33\% of these objects belong to galactic thin, thick, and thin/thick disks, respectively. In addition to the space velocity components, we derive the radial peculiar velocity (Vp) following the method given by \citet{Quireza07} and applied in Paper I and Paper II. Moreover, we calculate the vertical galactic height (Z) for the objects in the galactic thin and thick disks. Table \ref{Table3} shows the space velocity components, total space velocity, radial peculiar velocity, and galactic vertical height, as well as their related uncertainties, for only 46 verified objects that have the required data. A summary of these parameters, for the objects that located in the galactic thin and thick disks, is given in Table \ref{Table4}.The results indicate that the vertical heights of both populations are greater than those reported in Paper II, while the space velocity and rotational velocity component are comparable to those reported in Paper II (taking into account that the rotational solar velocity assumed here is 185 km/s versus 220 km/s in Paper II). The mean radial peculiar velocity of the galactic thin and thick disk is larger and smaller than in Paper II.1

\section{PNe that mimic the structure of IPNe}
Some mechanisms may modify the nebula's axisymmetry and generate symptoms that resemble those IPNe, providing the erroneous impression that the nebula is truly interacting with the ISM. \citet{Soker&Rappaport01} proposed that there are four major mechanisms that can lead to a large-scale deviation from axisymmetry: interaction with the ISM, local mass-loss events, a wide binary companion, and a close binary companion in an eccentric orbit. The presence of an enhancement in a portion of the nebular shell, PN asymmetry, and shift in CS location could all be attributed to other effects rather than interaction with the ISM.

The binarity of the PNe CS precursors may cause axisymmetry deviation, particularly in the inner parts of the PN.
\begin{enumerate}
\item {(Wide binary CS): According to \citet{Soker94}, very wide binaries with orbital periods comparable to or longer than the mass-loss episode will produce axisymmetry deviations. The asymmetry appears as an arc on the nebular outer edge when viewed from the pole. NGC 7662 is an example of PN with a structure that could be clarified by the presence of a wide binary system. This PN central star is displaced SE toward the rim, with bright knots settled in the same direction.}

\item {(Close binary CS): Assuming the CS is a close binary with an eccentric orbit, the tidally enhanced mass-loss rate will lead to a periodic variation in the mass-loss rate as well as a displacement of the nebula relative to the binary center of mass.}

\item {(Intermediate range binary CS): In the intermediate range, binary systems that have orbital periods in the range of several hundred to a few hundred thousand years may cause the PN to have a large-scale departure from axisymmetric structure \citep{Soker99}.}

\item {(A tertiary CS): A tertiary star may influence the mass-loss process in such a way that the descendant PN deviates from any kind of symmetry \citep{Bear&Soker17}.}
\end{enumerate}

Table 5 shows the mimic PN-ISM interacting objects. These objects are suggested to interact with ISM, while in fact they are not. The CSs of these nebulae move in different directions than those expected as interacting objects. Figure 7, shows a few examples of mimic IPNe. The majority of these objects have binary central stars. We present some examples of PNe that have binary CSs that impact their morphologies to resemble those interacting with the ISM. In all of these examples and others, the PA of their CS proper motions refer to inconsistent direction with that predicted by IPNe.

\begin{itemize}
\item {MWP\,1 (RX J2117+3412): The CS star of this nebula is a pulsating white dwarf of the GW Vir class. \citet{Corsico21} computed a rotation period of 1.04 days, from the TESS frequency spectrum, for the CS of the PN.}

\item {Jacoby\,1: It is a very circular nebula, with a general enhancement toward the rim that is particularly marked in the south direction. Due to this sign, \citet{Tweedy96} classified the object as IPN. The CS motion towards the north confirms that this object mimics the IPN. The CS of this object is classified as a close binary star by \citet{Douchin15} and as a pulsating star by \citet{Sowicka21}.}

\item {NGC 6853: The object is classified as IPN by \citet{Wareing07} due to the existence of a faint bow shock structure to the NW, which conflicts with the CS motion to NE. It has a bipolar shape and a medium mean radius of 0.368 pc ($\sim76,000$ AU). The CS of this PN has two wide binary companions, one of them with a projected separation of 2,453 AU from the CS and the other with a projected separation of 3,322 AU \citep{Gonzalez21}.}

\item {PN\,M\,1-46: The suggested interacting region of this nebula is to the W and NW \citep{Corradi03}. The proper CS motion indicates nebular motion towards the NE, which is inconsistent with the interacting region in the AGB halo. The CS of this PN was classified as variable star by \citet{Handler03}.}

\item {NGC\,2867: The core star of this PN was reported as a pulsating star by \citet{Gonzalez06} with a period of 769 s. Moreover, \citet{Aller20} found that the CS has significant periodic variability at 4.5 days.}

\item {PN\,WeDe\,1: \citet{DeMarco13} reported that the CS (of spectral type DA) of this nebula has a faint wide companion of spectral type later than the M5V spectral type. Further, \citet{Reindl21} assigned this star as a variable normal hot white dwarf with a period of 0.57 days and magnitude of 17.5.}

\item {PN K\,2-2: The CS of the object was classified as a probable close binary based on its radial velocity variability. Furthermore, the photometric study reported by \citet{DeMarco13} indicates that the spectral type of the CS companion is later than M1V.}

\item {PN\,IsWe\,1: The spectral type of the CS of this PN was defined as PG1159. \citet{DeMarco13} suggested that this CS has a companion of spectral type later than the M4V spectral type.}

\item {PN\,DeHt\,5: All of the evidence mentioned by \citet[\& references therein]{DeMarco13} shows that this object is an H II region instead of PNe, where it is ionized by the WD 2218+706 star that has a detected J-band excess, consistent with an M3V companion.}
\end{itemize}

\section{Conclusions}
In the present work, we verified the interaction between planetary nebulae and interstellar medium for 68 PNe. This verification is based mainly on the CS proper motion. We benefit from the recent precise proper motions and parallax measurements offered by the early phase of the third release of the Gaia space mission. Moreover, we reported the misclassification of 33 nebulae as PN-ISM interacting objects. The majority of the objects that mimic the structure of IPNe host binary CSs. Our analysis of the verified objects shows that 41\% and 41\% of them are members of galactic thin and thick disks, respectively. According to the classification of the IPNe suggested by \citet{Wareing07},
28\%, 37\%, 28\%, and 7\% belong to stages 1, 2, 3, and \textbf{3/4}, respectively. The kinematical results indicate that the mean galactic vertical height, space velocity, peculiar velocity, and galactic rotational velocity of the objects belonging to the galactic thin disk are $263\pm23$ pc, $35\pm7$ km/s, $32\pm6$ km/s, and $180\pm75$ km/s, respectively, and those of the galactic-thick disk are $600\pm90$ pc, $111\pm13$ km/s, $46\pm9$ km/s, and $158\pm34$ km/s, respectively.

\acknowledgments
This work was funded by the University of Jeddah, Jeddah, Saudi Arabia, under Grant No. (UJ-21-DR-10). The authors, therefore, acknowledge with thanks the University of Jeddah technical and financial support. This work has made use of data from the European Space Agency (ESA) mission Gaia, processed by the Gaia Data Processing and Analysis Consortium(DPAC). This research has made use of the SIMBAD database, operated at CDS, Strasbourg, France. This research has made use of "Aladin Sky Atals" developed at CDS, Strasbourg Observatory, France.

\begin{authorcontribution}
The contributions of the first, second, third, and fourth authors in this work are 40\%, 40\%, 10\%, and 10\%, respectively.
\end{authorcontribution}

\begin{fundinginformation}
This work was funded by the University of Jeddah, under Grant No. (UJ-21-DR-10).
\end{fundinginformation}

\begin{materialsavailability}
As, we mentioned in Section 2, all the observations that used in this study are collected from the Gaia, SIMBAD, and Aladin Sky Atals databases.
\end{materialsavailability}

\begin{ethics}
\begin{conflict}
Not applicable.
\end{conflict}
\end{ethics}


\begin{figure*}[t]
\includegraphics[width=17.0cm, height=14.0cm]{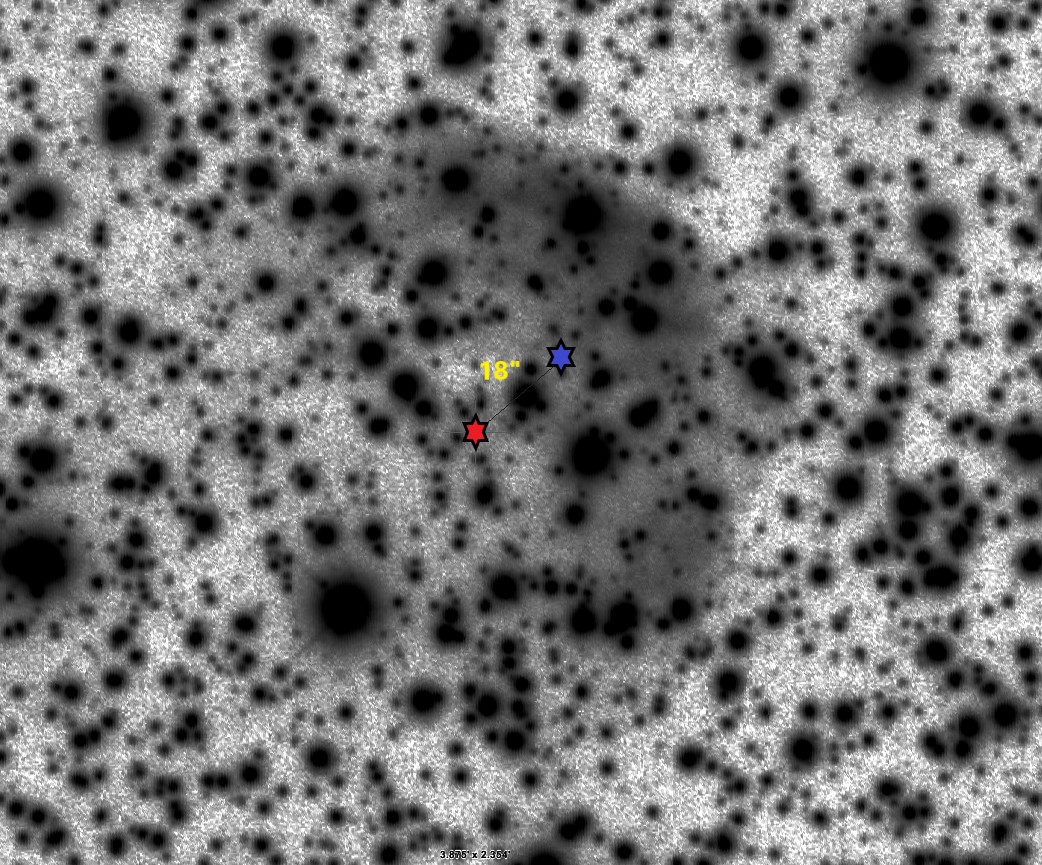}
\caption{Example of CS displacement from the nebular center as a result of the interaction with ISM. The blue star indicates the ionizing CS of the PN KFR\,1, which is shifted a distance of 18 arc-seconds from the PN center (red star). The CS shift indicates a rough movement in the same direction of the CS proper motion towards the nebular enhanced brightness region} \label{Figure1}
\end{figure*}

\begin{figure*}[t]
\includegraphics[width=17.0cm, height=24.0cm]{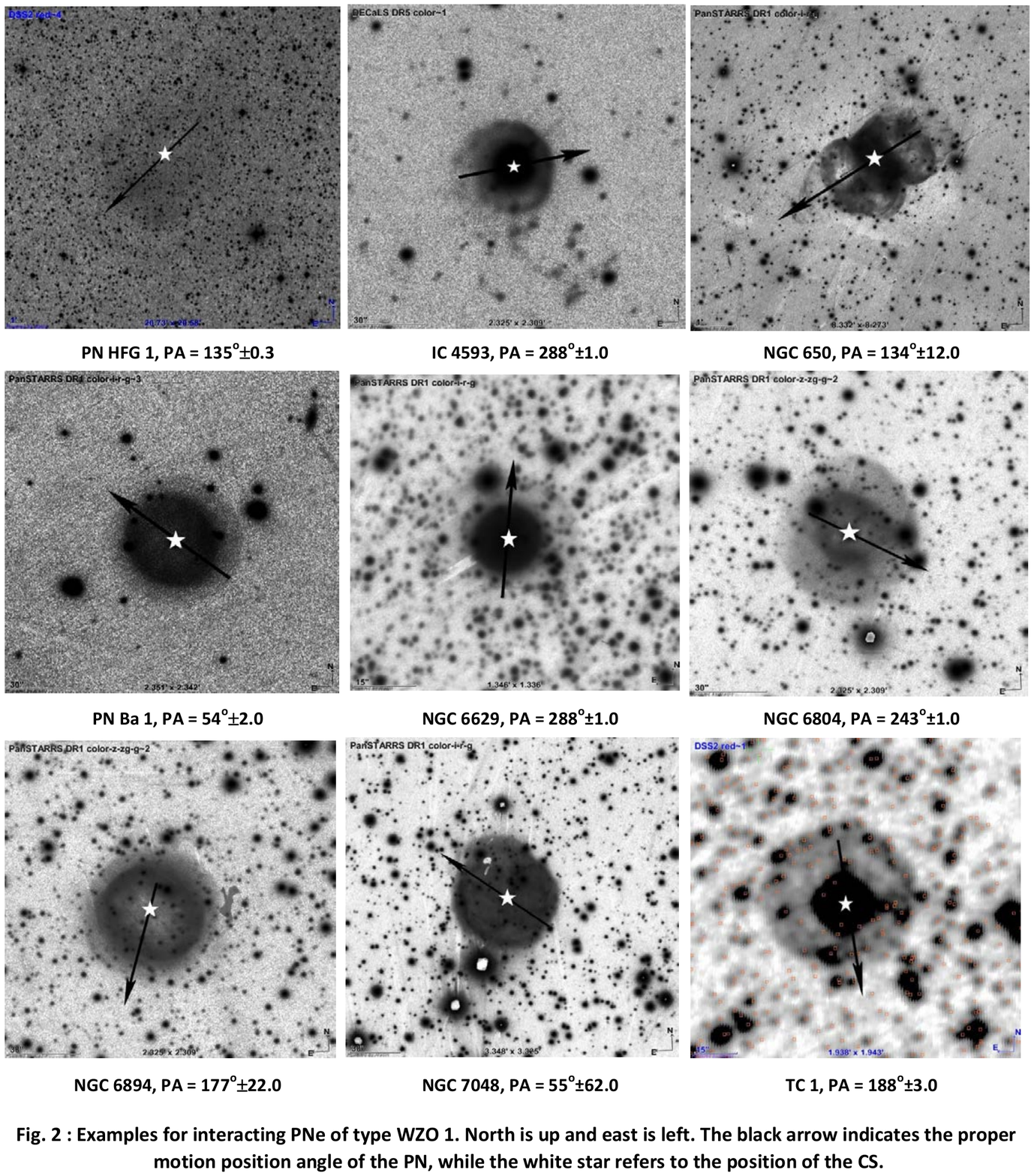}
 \label{Figure1}
\end{figure*}

\begin{figure*}[t]
\includegraphics[width=17.0cm, height=24.0cm]{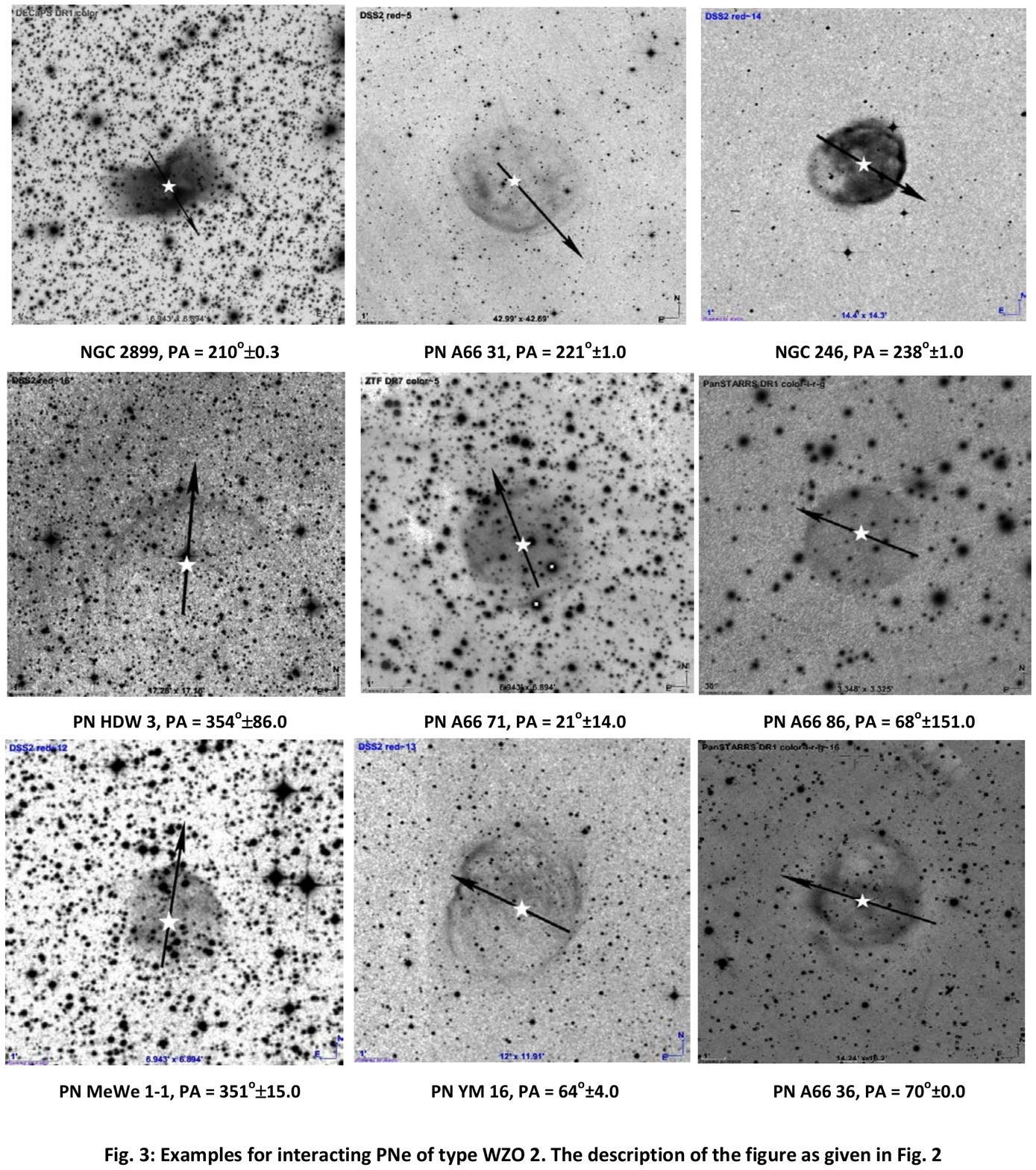}
 \label{Figure1}
\end{figure*}

\begin{figure*}[t]
\includegraphics[width=17.0cm, height=24.0cm]{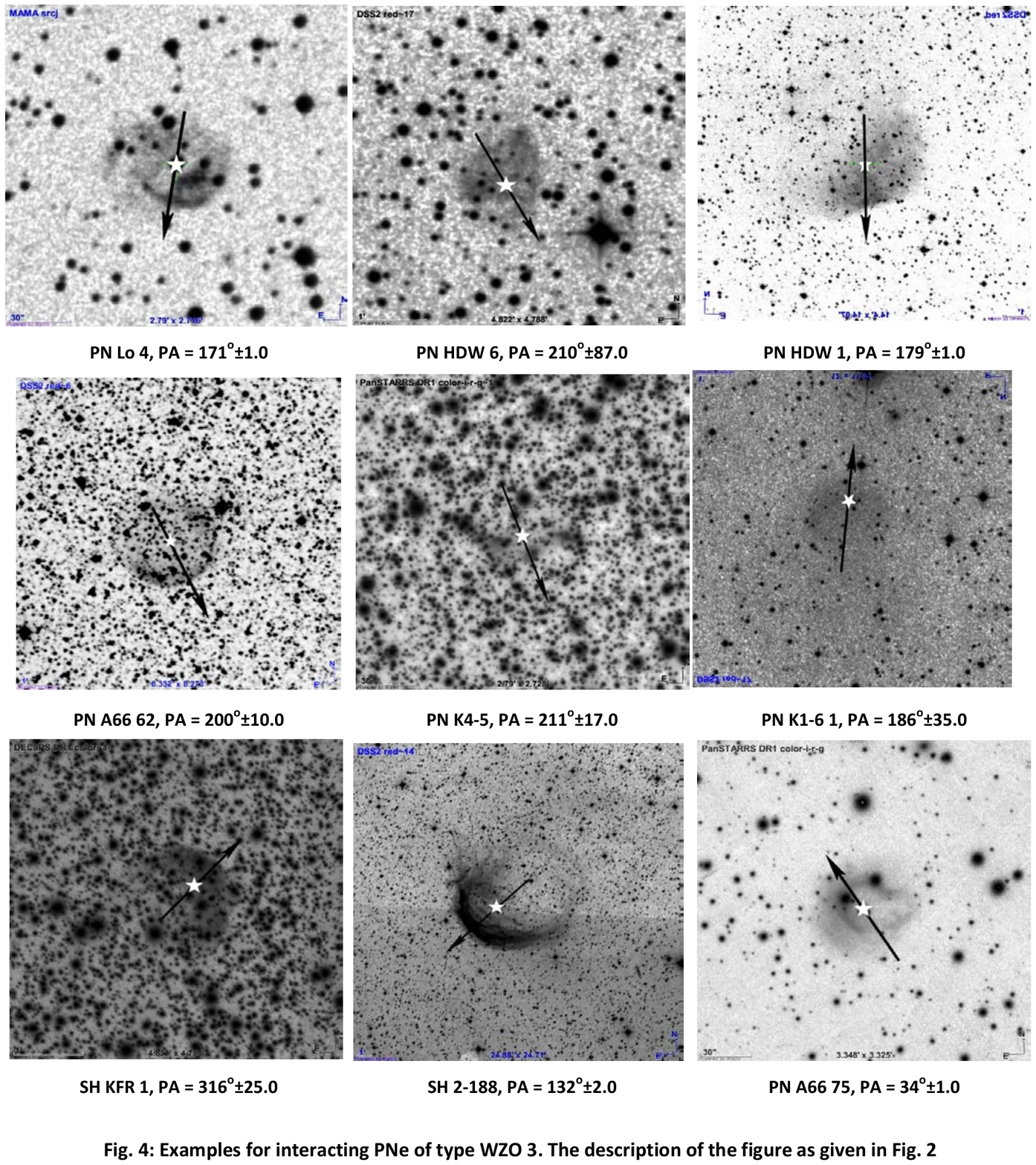}
 \label{Figure1}
\end{figure*}

\begin{figure*}[t]
\includegraphics[width=17.0cm, height=24.0cm]{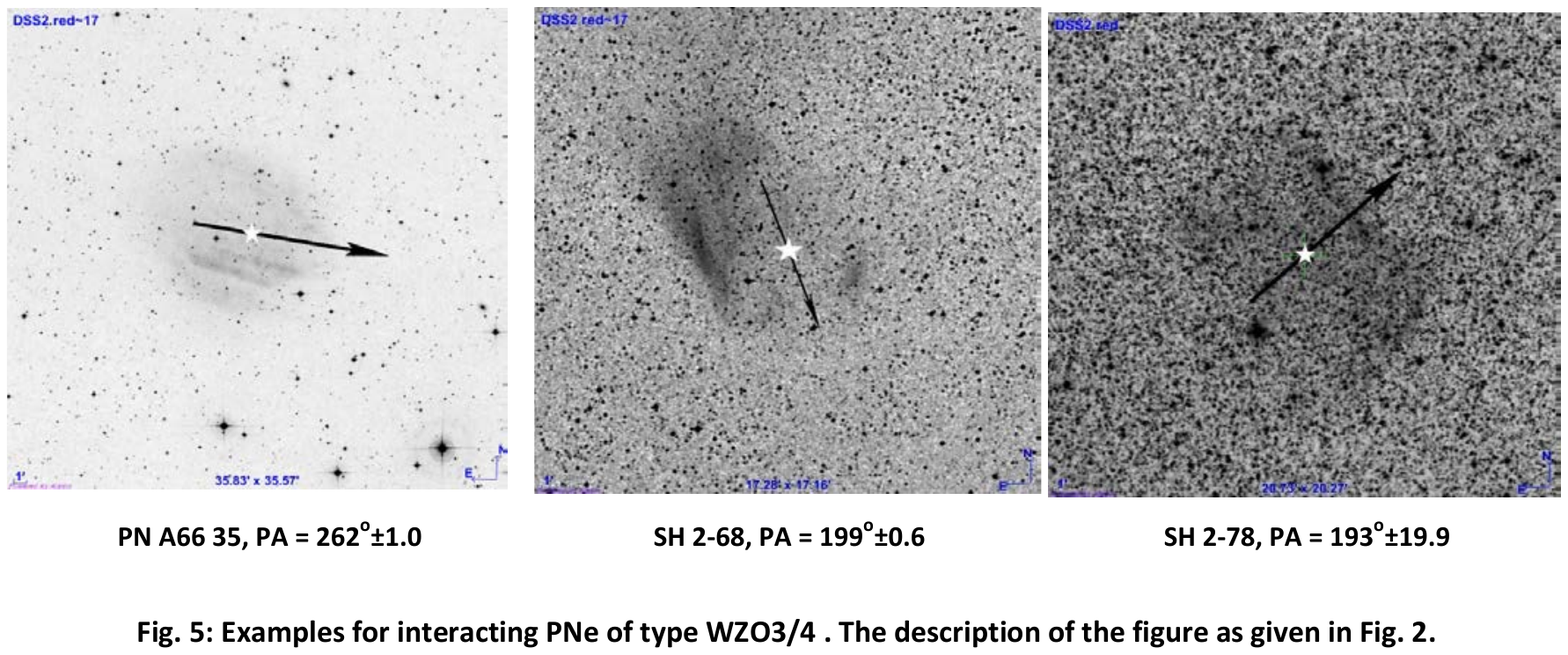}
 \label{Figure1}
\end{figure*}

\setcounter{figure}{5}

\begin{figure*}[t]
\includegraphics[width=17.0cm, height=12.0cm]{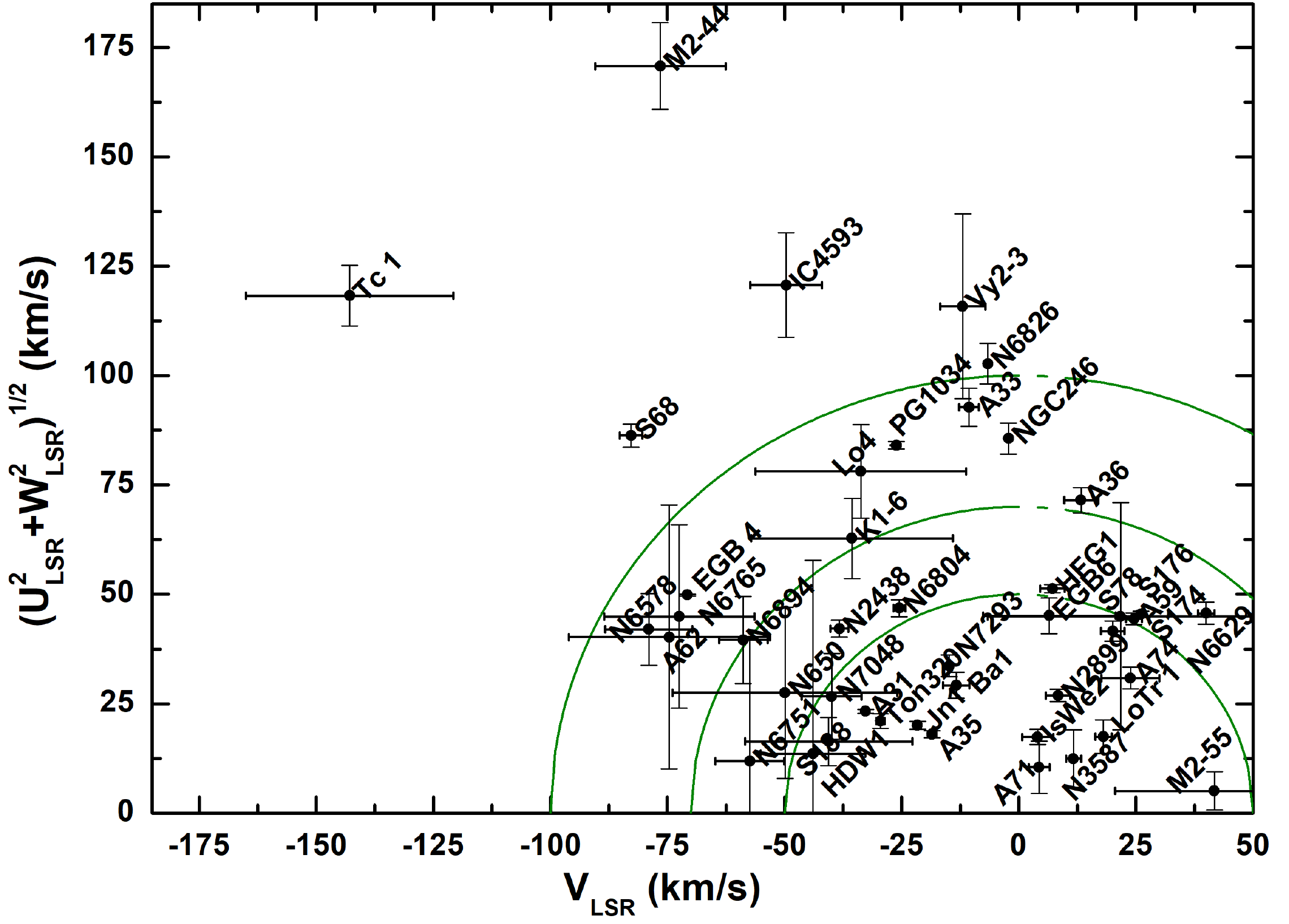}
\caption{Toomre diagram for the sample of verified IPNe. The objects belonging to galactic thin and thick disks are those with $V_S \leq 50\, \rm{km\,s^{-1}}$ (inside the smallest circle) and $200 \geq V_S \geq 70\, \rm{km\,s^{-1}}$ (outside the middle circle), respectively. The PNe lies between $70 \geq V_S \geq 50\, \rm{km\,s^{-1}}$ and are probable thin or thick disk objects} \label{Figure6}
\end{figure*}

\begin{figure*}[t]
\includegraphics[width=17.0cm, height=24.0cm]{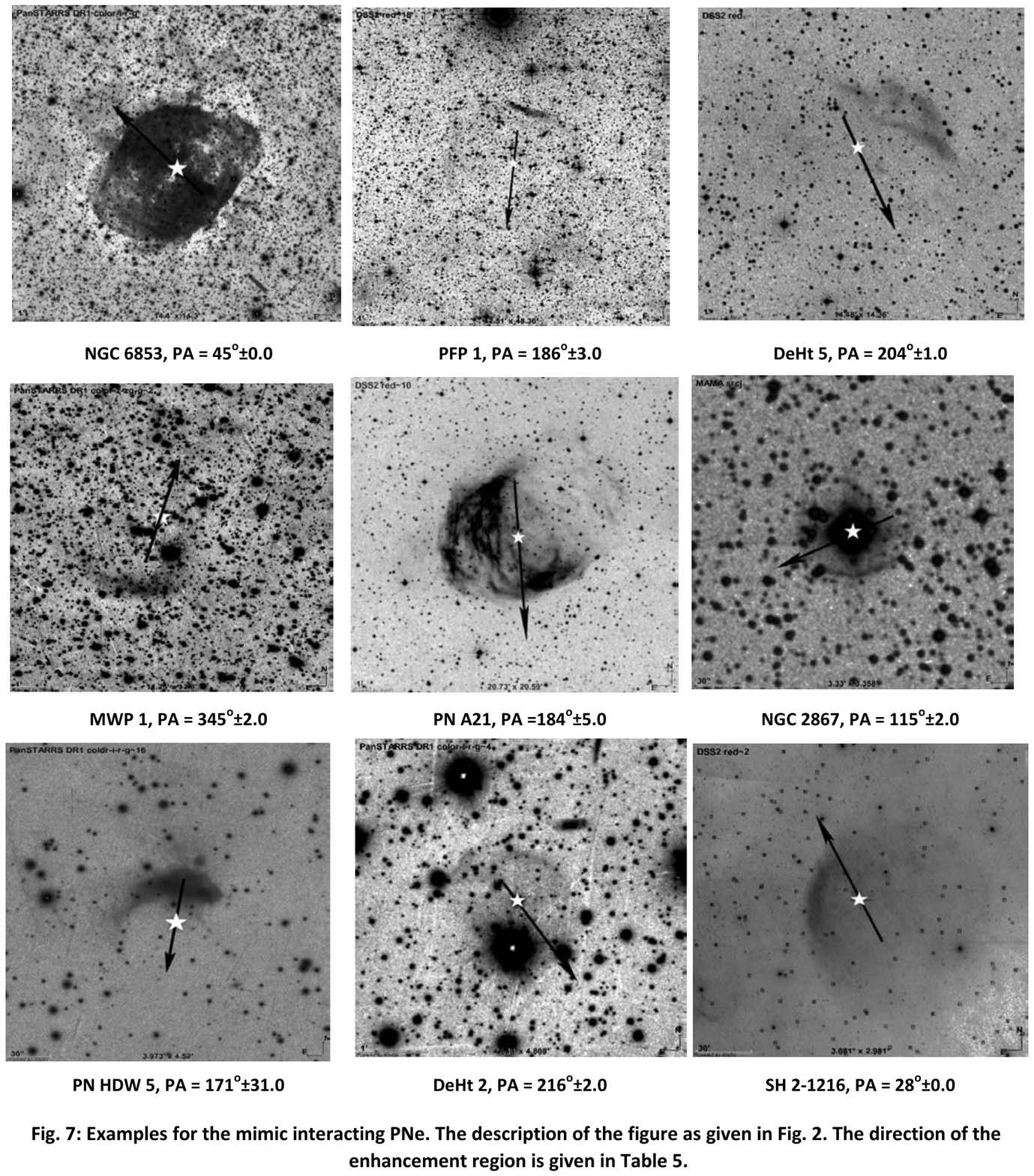}
 \label{Figure1}
\end{figure*}

\begin{table*}
\centering
\caption{The fundamental parameters of the verified IPNe group} \label{Table1}
\resizebox{\textwidth}{!}{
\begin{tabular}{lllccccccccccccc}
\hline

\# & PN	&	Gaia EDR3   	&	\multicolumn{5}{c}{Galactic and equatorial coordinates}		&		$\pi_{\rm Gaia}$	&	$\mu_{\alpha}$ &	$\mu_{\delta}$ 		&	G\,mag	&		Radius 	&	V$_{r}$		&	V$_{exp}$	\\\cline{4-8}
 &	&	designation	& l		& b		&	&	$\alpha$	&	$\delta$	&	&	mas	&	mas/yr	&	mas/yr		&			arcsec		&	kms$^{-1}$	&	kms$^{-1}$	\\
\hline
1	&	NGC\,6629	&	4089517157442187008	&	9.41	&	-5.05	&	&	276.43	&	-23.20	&	0.47	$\pm$	0.02	&	-1.09	$\pm$	0.03	&	3.94	$\pm$	0.02	&	12.7	&	8	$\pm$	1	&	14.6	$\pm$	1.3	&	20.2	\\
2	&	NGC\,6578	&	4094354493205707392	&	10.82	&	-1.83	&	&	274.07	&	-20.45	&	0.35	$\pm$	0.04	&	-2.26	$\pm$	0.04	&	-6.32	$\pm$	0.03	&	15.4	&	6	$\pm$	1	&	4.4	$\pm$	1.8	&	19.2	\\
3	&	IC\,4593	&	4457218245479455744	&	25.33	&	40.84	&	&	242.94	&	12.07	&	0.36	$\pm$	0.05	&	-9.23	$\pm$	0.04	&	1.30	$\pm$	0.04	&	11.2	&	8	$\pm$	1	&	22.0	$\pm$	0.5	&	12.3	\\
4	&	PN\, K\,4-5	&	4253203278927003648	&	26.65	&	-1.58	&	&	281.40	&	-6.31	&	0.48	$\pm$	0.23	&	-2.66	$\pm$	0.23	&	-6.03	$\pm$	0.17	&	18.5	&	3	$\pm$	0	&				&	20.0	\\
5	&	PN\,M\, 2-44	&	4259355871048895232	&	28.60	&	1.66	&	&	279.40	&	-3.10	&				&	-4.19	$\pm$	0.31	&	-6.61	$\pm$	0.27	&	18.7	&	5	$\pm$	1	&	106.0	$\pm$	40.0	&	12.0	\\
6	&	NGC\, 6751	&	4206136209740612352	&	29.23	&	-5.94	&	&	286.48	&	-5.99	&	0.27	$\pm$	0.04	&	-0.91	$\pm$	0.04	&	-3.14	$\pm$	0.04	&	13.7	&	12	$\pm$	1	&	-38.9	$\pm$	1.7	&	39.0	\\
7	&	SH\,2- 68	&	4276328581046447104	&	30.67	&	6.28	&	&	276.24	&	0.86	&	2.43	$\pm$	0.06	&	-19.08	$\pm$	0.06	&	-56.34	$\pm$	0.05	&	16.4	&	200	$\pm$	20	&	-7.0	$\pm$	3.5	&	6.6	\\
8	&	NGC\,7293	&	6628874205642084224	&	36.16	&	-57.12	&	&	337.41	&	-20.84	&	5.00	$\pm$	0.04	&	38.79	$\pm$	0.05	&	-3.35	$\pm$	0.04	&	13.5	&	426	$\pm$	43	&	-28.2	$\pm$	3.0	&	21.0	\\
9	&	PN\,YM\,16	&	4282704172233945216	&	38.71	&	1.97	&	&	283.74	&	6.04	&	1.13	$\pm$	0.79	&	6.35	$\pm$	0.78	&	-0.85	$\pm$	0.63	&	20.2	&	165	$\pm$	17	&				&	25.5	\\
10	&	IPHASX\,J185322.1+083018	&	4310333903026260608	&	40.72	&	3.44	&	&	283.34	&	8.50	&	0.61	$\pm$	0.07	&	-28.21	$\pm$	0.08	&	-31.96	$\pm$	0.07	&	16.8	&	55	$\pm$	6	&				&	20.0	\\
11	&	IPHASX\,J190454.0+101801	&	4312479187677796992	&	43.63	&	1.73	&	&	286.22	&	10.30	&	1.16	$\pm$	0.03	&	5.29	$\pm$	0.03	&	-36.38	$\pm$	0.03	&	14.6	&	9	$\pm$	1	&				&	20.0	\\
12	&	NGC\,6804	&	4296362443149857920	&	45.75	&	-4.59	&	&	292.90	&	9.23	&	1.40	$\pm$	0.07	&	-10.05	$\pm$	0.06	&	-9.35	$\pm$	0.06	&	14.0	&	27	$\pm$	3	&	-12.0	$\pm$	3.8	&	24.0	\\
13	&	SH\,2-78	&	4506484097383382272	&	46.83	&	3.84	&	&	285.79	&	14.12	&	1.41	$\pm$	0.10	&	-0.94	$\pm$	0.10	&	-3.87	$\pm$	0.11	&	17.6	&	298	$\pm$	30	&	31.7	$\pm$	2.3	&	20.0	\\
14	&	PN\,A66\,62	&	4314617943916816384	&	47.18	&	-4.29	&	&	293.32	&	10.62	&				&	-3.70	$\pm$	0.18	&	-9.76	$\pm$	0.18	&	18.6	&	81	$\pm$	8	&	-37.9	$\pm$	23.7	&	17.0	\\
15	&	PN\,A66\,52	&	4514604677947726976	&	50.41	&	5.29	&	&	286.13	&	17.95	&	0.32	$\pm$	0.11	&	0.38	$\pm$	0.08	&	-2.24	$\pm$	0.10	&	17.7	&	19	$\pm$	2	&				&	20.0	\\
16	&	IPHASX\,J194727.6+230816	&	1827892485806779776	&	59.76	&	-1.08	&	&	296.86	&	23.14	&				&	-3.16	$\pm$	0.18	&	-6.57	$\pm$	0.26	&	19.2	&				&				&	20.0	\\
17	&	NGC\,6765	&	2039515046435901440	&	62.46	&	9.56	&	&	287.78	&	30.55	&	0.28	$\pm$	0.08	&	-0.56	$\pm$	0.08	&	-3.72	$\pm$	0.08	&	17.6	&	17	$\pm$	2	&	-60.0	$\pm$	17.9	&	35.0	\\
18	&	IPHASX\,J194645.3+262211	&	 2027773464853061888	&	62.48	&	0.68	&	&	296.70	&	26.37	&	1.81	$\pm$	0.02	&	7.08	$\pm$	0.02	&	-3.74	$\pm$	0.02	&	14.8	&	145	$\pm$	15	&				&	20.0	\\
19	&	IPHASX\,J194240.6+275109	&	2031165660089336192	&	63.30	&	2.21	&	&	295.67	&	27.85	&				&	-2.27	$\pm$	0.83	&	-5.77	$\pm$	1.30	&	20.8	&	37	$\pm$	4	&				&	20.0	\\
20	&	NGC\,6894	&	 2053683628140770432	&	69.48	&	-2.62	&	&	304.10	&	30.57	&	0.36	$\pm$	0.34	&	-2.34	$\pm$	0.28	&	-4.77	$\pm$	0.34	&	19.7	&	27	$\pm$	3	&	-58.0	$\pm$	6.0	&	43.0	\\
21	&	PN\,A66\,74	&	1840395547924993152	&	72.66	&	-17.15	&	&	319.22	&	24.15	&	1.50	$\pm$	0.09	&	-3.92	$\pm$	0.09	&	-2.64	$\pm$	0.06	&	17.0	&	401	$\pm$	40	&	18.0	$\pm$	4.0	&	26.5	\\
22	&	NGC\,6826	&	2135352396915239808	&	83.56	&	12.79	&	&	296.20	&	50.53	&	0.75	$\pm$	0.04	&	-9.20	$\pm$	0.04	&	-11.42	$\pm$	0.05	&	10.6	&	13	$\pm$	1	&	-6.2	$\pm$	0.6	&	26.5	\\
23	&	PN\,A66\,71	&	2071605220993676544	&	85.00	&	4.49	&	&	308.10	&	47.35	&	1.01	$\pm$	0.21	&	0.34	$\pm$	0.22	&	0.09	$\pm$	0.21	&	19.2	&	79	$\pm$	8	&	-8.0	$\pm$	3.5	&	20.0	\\
24	&	NGC\,7048	&	2164192930525717248	&	88.78	&	-1.68	&	&	318.56	&	46.29	&	0.63	$\pm$	0.21	&	-0.75	$\pm$	0.24	&	-2.47	$\pm$	0.25	&	19.0	&	31	$\pm$	3	&	-50.2	$\pm$	8.1	&	15.0	\\
25	&	PN\, TK\,2	&	1634314740658456320	&	96.89	&	31.96	&	&	264.51	&	66.90	&	5.80	$\pm$	0.03	&	17.32	$\pm$	0.04	&	-16.42	$\pm$	0.04	&	14.5	&	1800	$\pm$	180	&				&	20.0	\\
26	&	PN\,A66\,75	&	2216998964996270208	&	101.86	&	8.75	&	&	321.60	&	62.89	&	0.28	$\pm$	0.05	&	-4.97	$\pm$	0.06	&	-2.03	$\pm$	0.05	&	17.1	&	29	$\pm$	3	&				&	20.0	\\
27	&	PN\,Jn\,1	&	2871119705335735552	&	104.21	&	-29.64	&	&	353.97	&	30.47	&	0.99	$\pm$	0.07	&	3.07	$\pm$	0.08	&	0.88	$\pm$	0.05	&	16.0	&	163	$\pm$	16	&	-67.0	$\pm$	30.0	&	20.0	\\
28	&	PN\,K\,1-6	&	2288467186442571008	&	107.04	&	21.38	&	&	301.06	&	74.43	&	3.86	$\pm$	0.03	&	-0.20	$\pm$	0.04	&	-28.19	$\pm$	0.05	&	12.3	&	90	$\pm$	9	&	-47.4	$\pm$	10.7	&	15.0	\\
29	&	Vy\,2-3	&	1941633699525901440	&	107.60	&	-13.32	&	&	350.74	&	46.90	&	0.14	$\pm$	0.03	&	-2.66	$\pm$	0.03	&	-1.09	$\pm$	0.03	&	14.6	&	2	$\pm$	0	&	-49.5	$\pm$	3.8	&	20.0	\\
30	&	PN\,IsWe\,2	&	2218688261534278912	&	107.73	&	7.81	&	&	333.34	&	65.90	&	1.20	$\pm$	0.10	&	0.31	$\pm$	0.12	&	-3.18	$\pm$	0.10	&	18.1	&	433	$\pm$	43	&	-8.0	$\pm$	2.0	&	20.0	\\
31	&	PN\,M\,2-55	&	2215132685747022464	&	116.30	&	8.53	&	&	352.97	&	70.37	&	1.52	$\pm$	0.05	&	-2.36	$\pm$	0.07	&	-3.72	$\pm$	0.07	&	16.9	&	25	$\pm$	2	&	22.6	$\pm$	2.3	&	31.0	\\
32	&	PN\,A66\,86	&	530858099521703808	&	118.80	&	8.24	&	&	0.38	&	70.71	&	0.40	$\pm$	1.26	&	0.91	$\pm$	1.32	&	0.67	$\pm$	1.36	&	20.6	&	35	$\pm$	4	&				&	20.0	\\
33	&	NGC\,246	&	2376592910265354368	&	118.86	&	-74.71	&	&	11.76	&	-11.87	&	1.78	$\pm$	0.08	&	-16.91	$\pm$	0.10	&	-9.24	$\pm$	0.09	&	11.8	&	122	$\pm$	12	&	-46.0	$\pm$	4.5	&	20.0	\\
34	&	SH\,2-174	&	2286296686066976896	&	120.22	&	18.43	&	&	356.26	&	80.95	&	3.43	$\pm$	0.03	&	-27.62	$\pm$	0.04	&	2.90	$\pm$	0.04	&	14.5	&	375	$\pm$	38	&	5.9	$\pm$	0.7	&	39.5	\\
35	&	SH\,2-176	&	421836329014389248	&	120.29	&	-5.39	&	&	7.97	&	57.38	&	0.90	$\pm$	0.14	&	1.23	$\pm$	0.12	&	-2.38	$\pm$	0.16	&	18.5	&	315	$\pm$	32	&	-37	$\pm$	30	&	20.0	\\
36	&	PN\,HDW\,1	&	535357713421191168	&	124.06	&	10.72	&	&	16.78	&	73.56	&	3.10	$\pm$	0.04	&	18.86	$\pm$	0.05	&	-11.54	$\pm$	0.04	&	16.4	&	135	$\pm$	14	&	-39.0	$\pm$	25.0	&	20.0	\\
37	&	SH\,2-188	&	509206447837376128	&	128.05	&	-4.07	&	&	22.64	&	58.41	&	1.04	$\pm$	0.07	&	5.70	$\pm$	0.05	&	-4.18	$\pm$	0.06	&	17.4	&	328	$\pm$	33	&	-26.0	$\pm$	10.0	&	20.0	\\
38	&	NGC\,650	&	406328443354164480	&	130.93	&	-10.50	&	&	25.58	&	51.58	&	0.29	$\pm$	0.20	&	2.59	$\pm$	0.23	&	-2.96	$\pm$	0.16	&	17.4	&	70	$\pm$	7	&	-19.1	$\pm$	1.2	&	40.0	\\
39	&	PN\,HFG\,1	&	468033345145186816	&	136.38	&	5.55	&	&	45.95	&	64.91	&	1.40	$\pm$	0.01	&	7.95	$\pm$	0.01	&	-4.88	$\pm$	0.02	&	14.0	&	240	$\pm$	24	&	-26.0	$\pm$	1.0	&	20.0	\\
40	&	SH\,2-200	&	466746538582572032	&	138.13	&	4.12	&	&	47.75	&	62.80	&	1.20	$\pm$	0.01	&	7.56	$\pm$	0.01	&	-7.16	$\pm$	0.01	&	12.5	&	176	$\pm$	18	&				&	20.0	\\
41	&	EGB\,4	&	1112772429499375488	&	143.60	&	23.82	&	&	97.39	&	71.08	&	2.66	$\pm$	0.02	&	-2.29	$\pm$	0.01	&	-28.81	$\pm$	0.02	&	13.0	&	56	$\pm$	6	&	-74.0	$\pm$	7.4	&	20.0	\\
42	&	NGC\,3587	&	843950873117830528	&	148.49	&	57.05	&	&	168.70	&	55.02	&	1.22	$\pm$	0.05	&	0.08	$\pm$	0.04	&	1.21	$\pm$	0.04	&	15.7	&	103	$\pm$	10	&	6.0	$\pm$	3.1	&	37.7	\\
43	&	PN\,HDW\,3	&	241918950690107264	&	149.50	&	-9.28	&	&	51.81	&	45.41	&	1.09	$\pm$	0.07	&	0.35	$\pm$	0.08	&	6.78	$\pm$	0.07	&	17.1	&	270	$\pm$	27	&				&	20.0	\\
44	&	PN\,PuWe\,1	&	997854527884948992	&	158.92	&	17.86	&	&	94.89	&	55.61	&	2.52	$\pm$	0.06	&	-1.08	$\pm$	0.04	&	-21.33	$\pm$	0.04	&	15.5	&	605	$\pm$	61	&				&	20.0	\\
45	&	PN\, Ba\,1	&	50865027106484096	&	171.30	&	-25.81	&	&	58.40	&	19.49	&	0.74	$\pm$	0.17	&	2.84	$\pm$	0.18	&	-0.99	$\pm$	0.13	&	18.0	&	27	$\pm$	3	&	-17.0	$\pm$	0.1	&	35.5	\\
46	&	TK\,1	&	708990321934310784	&	191.40	&	33.08	&	&	126.77	&	31.50	&	1.80	$\pm$	0.06	&	-6.42	$\pm$	0.06	&	-12.53	$\pm$	0.04	&	15.7	&	1013	$\pm$	101	&	33.8	$\pm$	3.5	&	20.0	\\
47	&	PN\, HDW\,6	&	3378885135797096192	&	192.53	&	7.22	&	&	100.04	&	21.41	&	0.80	$\pm$	0.76	&	-1.98	$\pm$	0.85	&	-3.19	$\pm$	0.80	&	20.1	&	44	$\pm$	4	&				&	21.0	\\
48	&	PN\, A66\,31	&	597324024095840512	&	219.13	&	31.29	&	&	133.55	&	8.90	&	1.83	$\pm$	0.06	&	-9.97	$\pm$	0.06	&	-8.57	$\pm$	0.04	&	15.5	&	465	$\pm$	47	&	44.0	$\pm$	1.0	&	17.5	\\
49	&	PN\, A66\,59	&	5084896688945791232	&	220.36	&	-53.93	&	&	53.31	&	-25.87	&	2.50	$\pm$	0.07	&	1.78	$\pm$	0.04	&	26.93	$\pm$	0.06	&	11.2	&	172	$\pm$	17	&	41.8	$\pm$	4.0	&	20.0	\\
50	&	EGB\,6	&	 615161091995252864	&	221.59	&	46.36	&	&	148.25	&	13.74	&	1.31	$\pm$	0.15	&	-15.54	$\pm$	0.14	&	0.99	$\pm$	0.13	&	16.0	&	360	$\pm$	36	&	-8.0	$\pm$	3.5	&	38.5	\\
51	&	PN\, LoTr\,1	&	2917223739718975616	&	228.21	&	-22.14	&	&	88.78	&	-22.90	&	0.43	$\pm$	0.03	&	-1.43	$\pm$	0.02	&	4.53	$\pm$	0.03	&	15.6	&	71	$\pm$	7	&	19.3	$\pm$	2.0	&	22.0	\\
52	&	NGC\,2438	&	3029214290412318720	&	231.80	&	4.12	&	&	115.46	&	-14.74	&	1.37	$\pm$	0.22	&	-5.62	$\pm$	0.19	&	1.52	$\pm$	0.18	&	17.1	&	40	$\pm$	4	&	75.4	$\pm$	2.5	&	21.2	\\
53	&	PN\, A66\,15	&	2924036760440812160	&	233.50	&	-16.00	&	&	96.00	&	-25.00	&	0.17	$\pm$	0.03	&	3.64	$\pm$	0.04	&	5.06	$\pm$	0.05	&	15.9	&	18	$\pm$	2	&	35.0	$\pm$	3.0	&	25.0	\\
54	&	PN\, A66\,33	&	3827045525522912128	&	238.02	&	34.86	&	&	144.79	&	-2.81	&	1.00	$\pm$	0.06	&	-14.89	$\pm$	0.06	&	9.41	$\pm$	0.05	&	15.9	&	135	$\pm$	14	&	60.1	$\pm$	4.0	&	37.0	\\
55	&	PG\, 1034+001	&	3806885288337214848	&	247.55	&	47.75	&	&	159.27	&	-0.14	&	5.16	$\pm$	0.05	&	-87.11	$\pm$	0.07	&	28.45	$\pm$	0.08	&	13.2	&	10800.0	$\pm$	1080.0	&	50.8	$\pm$	1.3	&	20.0	\\
56	&	PN\, A66\,34	&	5690534730341025408	&	248.71	&	29.54	&	&	146.40	&	-13.17	&	0.83	$\pm$	0.07	&	3.21	$\pm$	0.06	&	-9.16	$\pm$	0.06	&	16.4	&	144	$\pm$	14	&				&	34.0	\\
57	&	PN\,MeWe\,1-1	&	5317493563454092288	&	272.41	&	-5.97	&	&	133.40	&	-54.09	&				&	-8.13	$\pm$	0.30	&	6.33	$\pm$	0.27	&	19.1	&	70	$\pm$	7	&				&	20.0	\\
58	&	PN\, Lo\,4	&	5414927915911816704	&	274.31	&	9.11	&	&	151.44	&	-44.36	&	0.31	$\pm$	0.05	&	-6.30	$\pm$	0.05	&	0.30	$\pm$	0.06	&	16.6	&	20	$\pm$	2	&	33.0	$\pm$	20.0	&	20.0	\\
59	&	NGC\,2899	&	5307241785737968256	&	277.15	&	-3.83	&	&	141.76	&	-56.11	&	0.51	$\pm$	0.03	&	-2.66	$\pm$	0.03	&	-2.39	$\pm$	0.03	&	15.7	&	32	$\pm$	3	&	3.4	$\pm$	2.8	&	20.0	\\
60	&	PN\,K\,1-27	&	4648434175230114048	&	286.88	&	-29.58	&	&	89.26	&	-75.67	&	0.24	$\pm$	0.03	&	1.14	$\pm$	0.04	&	6.11	$\pm$	0.05	&	16.0	&	27	$\pm$	3	&	75.0	$\pm$	75.0	&	20.0	\\
61	&	KFR\,1	&	6071221840994998144	&	296.39	&	3.14	&	&	180.07	&	-59.08	&	0.62	$\pm$	0.41	&	-7.65	$\pm$	0.44	&	2.39	$\pm$	0.41	&	20.0	&	45	$\pm$	5	&				&	20.0	\\
62	&	PN\,HaTr\,1	&	 5860462744884685184	&	299.49	&	-4.12	&	&	184.14	&	-66.76	&				&	-5.55	$\pm$	0.35	&	0.42	$\pm$	0.43	&	19.7	&	34	$\pm$	3	&				&	20.0	\\
63	&	PN\,A66\,35	&	3499149202247569536	&	303.57	&	40.00	&	&	193.39	&	-22.87	&	7.29	$\pm$	0.32	&	-56.02	$\pm$	0.41	&	-13.05	$\pm$	0.27	&	9.4	&	386	$\pm$	39	&	-6.6	$\pm$	3.8	&	4.1	\\
64	&	PN\,MeWe\,2-4	&	6090271154808162048	&	314.09	&	10.68	&	&	210.31	&	-50.67	&	0.85	$\pm$	0.11	&	-10.43	$\pm$	0.12	&	-0.33	$\pm$	0.10	&	17.9	&	197	$\pm$	20	&				&	20.0	\\
65	&	PHR\, J1510-6754	&	5800264758167595648	&	315.45	&	-8.49	&	&	227.61	&	-67.92	&				&	-5.42	$\pm$	0.07	&	-2.66	$\pm$	0.10	&	18.1	&	106	$\pm$	11	&				&	20.0	\\
66	&	PN\,MeWe\,1-4	&	5897337306860137984	&	315.99	&	8.24	&	&	214.37	&	-52.44	&				&	-8.15	$\pm$	1.01	&	-4.02	$\pm$	1.04	&	20.1	&	62	$\pm$	6	&				&	20.0	\\
67	&	PN\, A66\,36	&	 6292074655679874688	&	318.46	&	41.50	&	&	205.17	&	-19.88	&	2.37	$\pm$	0.06	&	19.36	$\pm$	0.07	&	4.76	$\pm$	0.05	&	11.5	&	191	$\pm$	19	&	36.8	$\pm$	3.3	&	38.0	\\
68	&	Tc\,1	&	5954912374289120896	&	345.24	&	-8.83	&	&	266.40	&	-46.09	&	0.25	$\pm$	0.03	&	-2.33	$\pm$	0.04	&	-9.30	$\pm$	0.03	&	11.3	&	6	$\pm$	1	&	-84.1	$\pm$	4.7	&	8.0	\\
\hline
\end{tabular}}
\end{table*}

\begin{table*}
\centering
\caption{The position angle of CS proper motion and the nebular size and dynamical age of IPNe} \label{Table2}
\resizebox{\textwidth}{!}{
\begin{tabular}{llllllll}
 PN	&	PA($^{\rm o}$) &	Direction &	Reference & D\,(pc) & R\,(pc) & t\,(yrs) & WZO class \\
\hline
NGC\,6629		&	350	$\pm$	8	&	N   	&	\citet{Corradi03}	&	2132	$\pm$	103	&	0.08	$\pm$	0.009	&	4.04E+03	$\pm$	4.48E+02	&	WZO 1	\\
NGC\,6578		&	202	$\pm$	3	&	S   	&	\citet{Corradi03}	&	2828	$\pm$	318	&	0.08	$\pm$	0.012	&	4.19E+03	$\pm$	6.30E+02	&	WZO 1	\\
IC\,4593		&	288	$\pm$	1	&	NW   	&	\citet{Wareing07}	&	2788	$\pm$	353	&	0.10	$\pm$	0.016	&	8.12E+03	$\pm$	1.31E+03	&	WZO 1  	\\
PN\,M\, 2-44		&	218	$\pm$	15	&	SW   	&	\citet{Wareing07}	&	4029	$\pm$	432	&	0.11	$\pm$	0.016	&	8.66E+03	$\pm$	1.27E+03	&	WZO 1  	\\
NGC\, 6751		&	205	$\pm$	10	&	SW	&	\citet{Clark10}	&	3653	$\pm$	597	&	0.21	$\pm$	0.040	&	5.27E+03	$\pm$	1.01E+03	&	WZO 1  	\\
NGC\,7293		&	87	$\pm$	0	&	NE   	&	\citet{Borkowski90}	&	200	$\pm$	2	&	0.41	$\pm$	0.042	&	1.93E+04	$\pm$	1.94E+03	&	WZO 1	\\
NGC\,6804		&	243	$\pm$	1	&	SW   	&	\citet{Guerrero98}	&	714	$\pm$	34	&	0.09	$\pm$	0.010	&	3.78E+03	$\pm$	4.20E+02	&	WZO 1	\\
NGC\,6765		&	287	$\pm$	42	&	W	&	\citet{Wareing07}	&	3629	$\pm$	1022	&	0.30	$\pm$	0.089	&	8.38E+03	$\pm$	2.51E+03	&	WZO 1 	\\
NGC\,6894		&	177	$\pm$	22	&	SE-S  	&	\citet{Soker-Zucker97}	&	1496	$\pm$	253	&	0.20	$\pm$	0.039	&	4.54E+03	$\pm$	8.90E+02	&	WZO 1	\\
NGC\,6826		&	212	$\pm$	1	&	SW   	&	\citet{Aryal09}	&	1326	$\pm$	63	&	0.08	$\pm$	0.009	&	3.03E+03	$\pm$	3.36E+02	&	WZO 1	\\
NGC\,7048		&	55	$\pm$	18	&	NE	&	\citet{Fang18}	&	1594	$\pm$	539	&	0.24	$\pm$	0.084	&	1.55E+04	$\pm$	5.48E+03	&	WZO 1	\\
PN\,M\,2-55		&	205	$\pm$	7	&	SW 	&	\citet{Hsia20}	&	659	$\pm$	24	&	0.08	$\pm$	0.008	&	2.48E+03	$\pm$	2.63E+02	&	WZO 1	\\
NGC\,650		&	134	$\pm$	12	&	SE	&	\citet{Ramos18}	&	1060	$\pm$	300	&	0.36	$\pm$	0.108	&	8.79E+03	$\pm$	2.64E+03	&	WZO 1	\\
PN\,HFG\,1		&	135	$\pm$	0	&	SE 	&	\citet{Xilouris96}	&	716	$\pm$	8	&	0.83	$\pm$	0.084	&	4.09E+04	$\pm$	4.11E+03	&	WZO 1	\\
SH\,2-200		&	150	$\pm$	0	&	SE	&	\citet{Borkowski90}	&	831	$\pm$	8	&	0.71	$\pm$	0.071	&	3.48E+04	$\pm$	3.50E+03	&	WZO 1	\\
NGC\,3587		&	31	$\pm$	14	&	NE   	&	\citet{Corradi03}	&	821	$\pm$	31	&	0.41	$\pm$	0.044	&	1.06E+04	$\pm$	1.14E+03	&	WZO 1	\\
PN\, Ba\,1		&	54	$\pm$	2	&	E	&	\citet{Wareing07}	&	1343	$\pm$	306	&	0.17	$\pm$	0.043	&	4.81E+03	$\pm$	1.20E+03	&	WZO 1	\\
NGC\,2438		&	289	$\pm$	7	&	W	&	\citet{Ramos09}	&	732	$\pm$	119	&	0.14	$\pm$	0.027	&	6.53E+03	$\pm$	1.24E+03	&	WZO 1	\\
Tc\,1		&	188	$\pm$	3	&	 SW   	&	\citet{Corradi03}	&	3997	$\pm$	549	&	0.12	$\pm$	0.021	&	1.49E+04	$\pm$	2.53E+03	&	WZO 1	\\
\bf{Median value}		&				&		&		&				&	\bf{0.17}	$\pm$	\bf{0.039}	&	\bf{8.12E+03}	$\pm$	\bf{1.24E+03}	&		\\ \hline
PN\,YM\,16		&	64	$\pm$	4	&	NE	&	\citet{Dgani98}	&	990	$\pm$	280	&	0.79	$\pm$	0.238	&	3.05E+04	$\pm$	9.14E+03	&	WZO 2	\\
IPHASX\,J185322.1+083018		&	224	$\pm$	1	&	SW 	&	\citet{Sabin08}	&	1630	$\pm$	198	&	0.43	$\pm$	0.068	&	2.13E+04	$\pm$	3.35E+03	&	WZO 2	\\
IPHASX\,J190454.0+101801		&	170	$\pm$	1	&	SE	&	\citet{Sabin21}	&	863	$\pm$	19	&	0.04	$\pm$	0.004	&	1.85E+03	$\pm$	1.89E+02	&	WZO 2	\\
PN\,A66\,52		&	52	$\pm$	11	&	NE	&	\citet{Wareing07}	&	3110	$\pm$	1104	&	0.28	$\pm$	0.103	&	1.37E+04	$\pm$	5.05E+03	&	WZO 2	\\
IPHASX\,J194727.6+230816		&	204	$\pm$	13	&	SW	&	\citet{Sabin21}	&				&				&				&	WZO 2	\\
IPHASX\,J194645.3+262211		&	79	$\pm$	0	&	SE	&	\citet{Sabin08}	&	554	$\pm$	7	&	0.39	$\pm$	0.039	&	1.91E+04	$\pm$	1.93E+03	&	WZO 2	\\
IPHASX\,J194240.6+275109		&	185	$\pm$	74	&	S	&	\citet{Sabin21}	&	6870	$\pm$	3990	&	1.23	$\pm$	0.726	&	6.04E+04	$\pm$	3.56E+04	&	WZO 2	\\
PN\,A66\,74		&	296	$\pm$	5	&	SW	&	\citet{Borkowski90}	&	665	$\pm$	38	&	1.29	$\pm$	0.149	&	4.79E+04	$\pm$	5.52E+03	&	WZO 2	\\
PN\,A66\,71		&	21	$\pm$	14	&	NE	&	\citet{Soker97}	&	990	$\pm$	205	&	0.38	$\pm$	0.087	&	1.85E+04	$\pm$	4.26E+03	&	WZO 2	\\
PN\,Jn\,1		&	77	$\pm$	1	&	NE	&	\citet{Wareing07}	&	1006	$\pm$	66	&	0.79	$\pm$	0.095	&	3.90E+04	$\pm$	4.66E+03	&	WZO 2 	\\
Vy\,2-3		&	85	$\pm$	1	&	NE	&	\citet{Guerrero98}	&	6923	$\pm$	1489	&	0.08	$\pm$	0.018	&	3.79E+03	$\pm$	8.98E+02	&	WZO 2	\\
PN\,A66\,86		&	68	$\pm$	151	&	NE   	&	\citet{Wareing07}	&	3100	$\pm$	870	&	0.53	$\pm$	0.157	&	2.58E+04	$\pm$	7.69E+03	&	WZO 2  	\\
NGC\,246		&	238	$\pm$	1	&	SW   	&	\citet{Borkowski90}	&	561	$\pm$	25	&	0.33	$\pm$	0.036	&	1.62E+04	$\pm$	1.78E+03	&	WZO 2	\\
PN\,HDW\,3		&	354	$\pm$	86	&	N-NE	&	\citet{Tweedy96}	&	916	$\pm$	59	&	1.20	$\pm$	0.143	&	5.88E+04	$\pm$	7.00E+03	&	WZO 2	\\
PN\,PuWe\,1		&	182	$\pm$	7	&	S	&	\citet{Tweedy96}	&	397	$\pm$	9	&	1.17	$\pm$	0.120	&	5.71E+04	$\pm$	5.87E+03	&	WZO 2	\\
PN\, A66\,31		&	221	$\pm$	1	&	S	&	\citet{Tweedy94b}	&	548	$\pm$	17	&	1.24	$\pm$	0.129	&	6.92E+04	$\pm$	7.23E+03	&	WZO 2	\\
EGB\,6		&	288	$\pm$	2	&	W   	&	\citet{Tweedy96}	&	761	$\pm$	85	&	1.33	$\pm$	0.199	&	3.39E+04	$\pm$	5.06E+03	&	WZO 2	\\
PN\, LoTr\,1		&	333	$\pm$	6	&	NW   	&	\citet{Bond89}	&	2350	$\pm$	174	&	0.81	$\pm$	0.101	&	3.61E+04	$\pm$	4.49E+03	&	WZO 2	\\
PN\, A66\,15		&	65	$\pm$	1	&	SE	&	\citet{Soker97}	&	6061	$\pm$	1269	&	0.52	$\pm$	0.122	&	2.06E+04	$\pm$	4.77E+03	&	WZO 2	\\
PN\, A66\,33		&	314	$\pm$	1	&	NW	&	\citet{Borkowski90}	&	995	$\pm$	58	&	0.65	$\pm$	0.076	&	1.73E+04	$\pm$	2.00E+03	&	WZO 2	\\
PN\, A66\,34		&	141	$\pm$	3	&	SE   	&	\citet{Tweedy94b}	&	1207	$\pm$	96	&	0.84	$\pm$	0.107	&	2.42E+04	$\pm$	3.10E+03	&	WZO 2	\\
PN\,MeWe\,1-1		&	351	$\pm$	15	&	NW   	&	\citet{Kerber2000}	&	2170	$\pm$	620	&	0.74	$\pm$	0.224	&	3.62E+04	$\pm$	1.10E+04	&	WZO 2	\\
NGC\,2899		&	210	$\pm$	3	&	SW   	&	\citet{Soker97}	&	1971	$\pm$	109	&	0.31	$\pm$	0.035	&	1.50E+04	$\pm$	1.72E+03	&	WZO 2	\\
PN\,K\,1-27		&	350	$\pm$	14	&	NW	&	\citet{Soker97}	&	4231	$\pm$	619	&	0.55	$\pm$	0.098	&	2.72E+04	$\pm$	4.81E+03	&	WZO 2	\\
PN\, A66\,36		&	70	$\pm$	0	&	E   	&	\citet{Borkowski90}	&	421	$\pm$	11	&	0.39	$\pm$	0.040	&	1.01E+04	$\pm$	1.04E+03	&	WZO 2	\\
\bf{Median value}		&				&		&		&				&	\bf{0.60}	$\pm$	\bf{0.10}	&	\bf{2.51E+04}	$\pm$	\bf{4.71E+03}	&		\\ \hline
PN\, K\,4-5		&	211	$\pm$	17	&	S	&	\citet{Wareing07}	&	2067	$\pm$	1002	&	0.03	$\pm$	0.014	&	1.38E+03	$\pm$	6.81E+02	&	WZO 3	\\
PN\,A66\,62		&	205	$\pm$	10	&	SW	&	\citet{Xilouris96}	&	1560	$\pm$	450	&	0.61	$\pm$	0.186	&	3.51E+04	$\pm$	1.07E+04	&	WZO 3	\\
PN\,A66\,75		&	34	$\pm$	1	&	NE-E	&	\citet{Tweedy96}	&	3604	$\pm$	674	&	0.50	$\pm$	0.106	&	2.44E+04	$\pm$	5.18E+03	&	WZO 3 	\\
PN\,K\,1-6		&	186	$\pm$	35	&	S   	&	\citet{Frew11}	&	259	$\pm$	2	&	0.11	$\pm$	0.011	&	7.34E+03	$\pm$	7.37E+02	&	WZO 3	\\
PN\,IsWe\,2		&	136	$\pm$	54	&	SE	&	\citet{Xilouris96}	&	830	$\pm$	66	&	1.74	$\pm$	0.222	&	8.54E+04	$\pm$	1.09E+04	&	WZO 3	\\
SH\,2-174		&	288	$\pm$	2	&	NW	&	\citet{Ransom15}	&	292	$\pm$	3	&	0.53	$\pm$	0.053	&	1.32E+04	$\pm$	1.32E+03	&	WZO 3	\\
SH\,2-176		&	123	$\pm$	14	&	SE   	&	\citet{Tweedy96}	&	1110	$\pm$	179	&	1.70	$\pm$	0.321	&	8.32E+04	$\pm$	1.58E+04	&	WZO 3	\\
PN\,HDW\,1		&	179	$\pm$	1	&	SE-S  	&	\citet{Soker97}	&	322	$\pm$	4	&	0.21	$\pm$	0.021	&	1.03E+04	$\pm$	1.04E+03	&	WZO 3	\\
SH\,2-188		&	132	$\pm$	2	&	SE   	&	\citet{Xilouris96}	&	959	$\pm$	67	&	1.52	$\pm$	0.186	&	7.48E+04	$\pm$	9.14E+03	&	WZO 3 	\\
EGB\,4		&	181	$\pm$	104	&	S   	&	\citet{Wareing07}	&	377	$\pm$	3	&	0.10	$\pm$	0.010	&	4.97E+03	$\pm$	4.98E+02	&	WZO 3	\\
TK\,1		&	196	$\pm$	2	&	SE 	&	\citet{Tweedy94a}	&	556	$\pm$	18	&	2.73	$\pm$	0.287	&	1.34E+05	$\pm$	1.41E+04	&	WZO 3	\\
PN\, HDW\,6		&	210	$\pm$	87	&	SW	&	\citet{Soker97}	&	4180	$\pm$	1280	&	0.89	$\pm$	0.286	&	4.14E+04	$\pm$	1.33E+04	&	WZO 3	\\
PN\, A66\,59		&	19	$\pm$	34	&	N	&	\citet{Wareing07}	&	548	$\pm$	17	&	1.24	$\pm$	0.129	&	6.92E+04	$\pm$	7.23E+03	&	WZO 2	\\
PN\, Lo\,4		&	171	$\pm$	1	&	S	&	\citet{Kerber2000}	&	3191	$\pm$	529	&	0.31	$\pm$	0.060	&	1.53E+04	$\pm$	2.95E+03	&	WZO 3	\\
KFR\,1		&	316	$\pm$	25	&	NW	&	\citet{Kerber2000}	&	2800	$\pm$	780	&	0.61	$\pm$	0.182	&	3.01E+04	$\pm$	8.92E+03	&	WZO 3	\\
PN\,HaTr\,1		&	82	$\pm$	11	&	E	&	\citet{Kerber2000}	&	2590	$\pm$	770	&	0.43	$\pm$	0.135	&	2.11E+04	$\pm$	6.62E+03	&	WZO 3	\\
PN\,MeWe\,2-4		&	319	$\pm$	3	&	NW   	&	\citet{Kerber2000}	&	1179	$\pm$	151	&	1.13	$\pm$	0.183	&	5.52E+04	$\pm$	8.99E+03	&	WZO 3	\\
PHR\, J1510-6754		&	162	$\pm$	5	&	SE	&	\citet{Sabin08}	&	1670	$\pm$	300	&	0.86	$\pm$	0.177	&	4.22E+04	$\pm$	8.67E+03	&	WZO 3	\\
PN\,MeWe\,1-4		&	209	$\pm$	33	&	SW   	&	\citet{Kerber2000}	&	2170	$\pm$	630	&	0.65	$\pm$	0.199	&	3.17E+04	$\pm$	9.74E+03	&	WZO 3	\\
\bf{Median value}		&				&		&		&				&	\bf{0.61}	$\pm$	\bf{0.18}	&	\bf{3.17E04}	$\pm$	\bf{8.67E+03}	&		\\  \ hline
SH\,2- 68		&	201	$\pm$	1	&	SE-S 	&	\citet{Wareing10}	&	412	$\pm$	10	&	0.40	$\pm$	0.041	&	5.93E+04	$\pm$	6.11E+03	&	WZO 3/4	\\
SH\,2-78		&	312	$\pm$	32	&	NW	&	\citet{Tweedy96}	&	707	$\pm$	52	&	1.02	$\pm$	0.126	&	5.00E+04	$\pm$	6.20E+03	&	WZO 3/4	\\
PN\, TK\,2		&	158	$\pm$	0	&	SE	&	\citet{Tweedy96}	&	172	$\pm$	1	&	1.50	$\pm$	0.151	&	7.37E+04	$\pm$	7.38E+03	&	WZO 3/4	\\
PG\, 1034+001		&	298	$\pm$	0	&	W	&	\citet{Tweedy94a}	&	194	$\pm$	2	&	10.14	$\pm$	1.019	&	4.97E+05	$\pm$	5.00E+04	&	WZO 3/4	\\
PN\,A66\,35		&	261	$\pm$	1	&	W   	&	\citet{Borkowski90}	&	137	$\pm$	6	&	0.26	$\pm$	0.028	&	6.20E+04	$\pm$	6.76E+03	&	WZO 3/4	\\
\bf{Median value}		&				&		&		&				&	\bf{1.02}	$\pm$	\bf{0.13}	&	\bf{6.20E04}	$\pm$	\bf{6.76E+03}	&		\\
\hline																								
Median value of the entire sample		&				&		&		&				&	\bf{0.43}	$\pm$	\bf{0.09}	&	\bf{2.13E04}	$\pm$	\bf{4.26E+03}	&		\\

\hline
\end{tabular}}
\end{table*}

\begin{table*}
\centering
\caption{The space velocity components of the verified IPNe} \label{Table3}
\begin{tabular}{lllllll}
\hline
 PN	        &	$U_{LSR}$ (km/s)          &	     $V_{LSR}$  (km/s)    &     $W_{LSR}$  (km/s)      &           Vs\,(km/s)     &   Vp\,(km/s) &   Z\,(pc)\\
\hline
NGC\,6629	&	31.8	$\pm$	3.2	&	225.0	$\pm$	10.0	&	32.7	$\pm$	1.8	&	60.7	$\pm$	2.2	&	22.3	$\pm$	1.9	&	188	$\pm$	9	\\
NGC\,6578	&	41.4	$\pm$	8.3	&	106.0	$\pm$	12.4	&	-7.1	$\pm$	0.8	&	89.5	$\pm$	9.0	&	12.6	$\pm$	5.4	&	90	$\pm$	10	\\
IC\,4593	&	-18.1	$\pm$	4.3	&	135.3	$\pm$	20.7	&	119.3	$\pm$	12.1	&	130.5	$\pm$	11.4	&	58.0	$\pm$	0.7	&	1823	$\pm$	231	\\
PN\,M\, 2-44	&	169.3	$\pm$	9.9	&	108.5	$\pm$	19.8	&	21.7	$\pm$	8.9	&	187.1	$\pm$	10.7	&	83.4	$\pm$	4.7	&	116	$\pm$	12	\\
NGC\, 6751	&	2.8	$\pm$	1.8	&	122.4	$\pm$	14.5	&	1.0	$\pm$	0.3	&	62.6	$\pm$	155.5	&	55.2	$\pm$	1.1	&	378	$\pm$	62	\\
SH\,2- 68	&	85.9	$\pm$	2.6	&	102.2	$\pm$	3.0	&	-8.6	$\pm$	0.3	&	119.6	$\pm$	2.5	&	15.6	$\pm$	1.9	&	45	$\pm$	1	\\
NGC\,7293	&	-31.5	$\pm$	1.0	&	170.2	$\pm$	8.1	&	11.8	$\pm$	6.5	&	36.8	$\pm$	2.3	&	46.2	$\pm$	2.7	&	168	$\pm$	1	\\
NGC\,6804	&	41.4	$\pm$	2.2	&	159.5	$\pm$	7.7	&	21.8	$\pm$	1.1	&	53.3	$\pm$	1.8	&	15.1	$\pm$	0.5	&	57	$\pm$	3	\\
SH\,2-78	&	41.2	$\pm$	2.3	&	205.1	$\pm$	25.7	&	5.9	$\pm$	1.9	&	46.2	$\pm$	2.3	&	33.6	$\pm$	2.8	&	47	$\pm$	3	\\
PN\,A66\,62	&	40.2	$\pm$	30.2	&	110.4	$\pm$	31.8	&	-0.7	$\pm$	0.4	&	84.8	$\pm$	23.7	&	47.5	$\pm$	15.4	&	116.8	$\pm$	33.7	\\
NGC\,6765	&	39.1	$\pm$	24.0	&	112.5	$\pm$	24.9	&	-22.2	$\pm$	4.8	&	85.3	$\pm$	17.6	&	66.8	$\pm$	13.1	&	603	$\pm$	170	\\
NGC\,6894	&	37.1	$\pm$	10.6	&	126.2	$\pm$	11.1	&	-13.7	$\pm$	2.6	&	70.9	$\pm$	7.0	&	73.0	$\pm$	4.6	&	68	$\pm$	12	\\
PN\,A66\,74	&	30.7	$\pm$	2.5	&	208.8	$\pm$	54.4	&	3.6	$\pm$	1.6	&	39.0	$\pm$	4.3	&	2.3	$\pm$	4.6	&	196	$\pm$	11	\\
NGC\,6826	&	99.7	$\pm$	4.8	&	178.4	$\pm$	9.9	&	24.5	$\pm$	1.2	&	102.9	$\pm$	4.6	&	12.1	$\pm$	0.2	&	294	$\pm$	14	\\
PN\,A66\,71	&	8.6	$\pm$	6.2	&	189.3	$\pm$	98.2	&	6.0	$\pm$	5.5	&	11.4	$\pm$	5.6	&	20.3	$\pm$	0.6	&	78	$\pm$	16	\\
NGC\,7048	&	26.7	$\pm$	9.8	&	145.0	$\pm$	23.1	&	-0.5	$\pm$	0.2	&	48.0	$\pm$	7.6	&	64.2	$\pm$	6.4	&	47	$\pm$	16	\\
PN\,Jn\,1	&	1.5	$\pm$	0.2	&	163.3	$\pm$	5.7	&	20.0	$\pm$	0.9	&	29.5	$\pm$	0.8	&	89.1	$\pm$	27.1	&	497	$\pm$	33	\\
PN\,K\,1-6	&	55.5	$\pm$	9.4	&	149.3	$\pm$	90.2	&	-29.3	$\pm$	8.2	&	72.2	$\pm$	13.3	&	56.4	$\pm$	8.6	&	94	$\pm$	1	\\
Vy\,2-3	&	114.9	$\pm$	21.3	&	173.0	$\pm$	69.6	&	14.1	$\pm$	3.1	&	116.4	$\pm$	21.1	&	4.0	$\pm$	3.4	&	1595	$\pm$	343	\\
PN\,IsWe\,2	&	16.8	$\pm$	1.8	&	189.0	$\pm$	157.2	&	-4.5	$\pm$	0.4	&	17.9	$\pm$	1.8	&	22.7	$\pm$	1.1	&	113	$\pm$	9	\\
PN\,M\,2-55	&	3.6	$\pm$	5.2	&	226.8	$\pm$	115.1	&	3.6	$\pm$	3.2	&	42.1	$\pm$	21.0	&	6.0	$\pm$	2.1	&	98	$\pm$	4	\\
NGC\,246	&	68.0	$\pm$	2.7	&	182.8	$\pm$	23.8	&	52.0	$\pm$	4.6	&	85.6	$\pm$	3.6	&	76.7	$\pm$	4.6	&	541	$\pm$	24	\\
SH\,2-174	&	39.5	$\pm$	0.7	&	211.1	$\pm$	7.8	&	22.6	$\pm$	0.5	&	52.5	$\pm$	0.6	&	14.2	$\pm$	0.8	&	92	$\pm$	1	\\
SH\,2-176	&	44.9	$\pm$	26.0	&	206.7	$\pm$	269.9	&	-2.0	$\pm$	1.5	&	50.0	$\pm$	26.4	&	47.8	$\pm$	27.4	&	104	$\pm$	17	\\
PN\,HDW\,1	&	5.1	$\pm$	14.4	&	144.4	$\pm$	63.5	&	-15.5	$\pm$	3.3	&	43.8	$\pm$	16.7	&	58.0	$\pm$	22.1	&	60	$\pm$	1	\\
SH\,2-188	&	12.7	$\pm$	47.3	&	141.1	$\pm$	35.9	&	-4.8	$\pm$	0.6	&	46.0	$\pm$	16.9	&	36.5	$\pm$	9.7	&	68	$\pm$	5	\\
NGC\,650	&	-8.6	$\pm$	10.8	&	135.1	$\pm$	64.8	&	-26.1	$\pm$	20.3	&	57.0	$\pm$	23.0	&	3.1	$\pm$	1.3	&	619	$\pm$	426	\\
PN\,HFG\,1	&	50.9	$\pm$	0.8	&	192.2	$\pm$	70.4	&	6.2	$\pm$	3.4	&	51.8	$\pm$	1.0	&	35.4	$\pm$	1.0	&	69	$\pm$	1	\\
EGB\,4	&	31.2	$\pm$	0.3	&	114.3	$\pm$	0.4	&	-38.9	$\pm$	0.1	&	86.5	$\pm$	0.2	&	74.0	$\pm$	6.7	&	152	$\pm$	1	\\
NGC\,3587	&	6.8	$\pm$	4.2	&	196.7	$\pm$	27.5	&	10.4	$\pm$	7.3	&	17.1	$\pm$	4.9	&	18.0	$\pm$	4.1	&	689	$\pm$	26	\\
PN\, Ba\,1	&	19.2	$\pm$	3.1	&	171.6	$\pm$	35.9	&	22.0	$\pm$	2.9	&	32.1	$\pm$	3.0	&	28.0	$\pm$	0.3	&	585	$\pm$	133	\\
Ton\,320 	&	-20.8	$\pm$	1.6	&	155.5	$\pm$	4.9	&	2.8	$\pm$	1.1	&	36.3	$\pm$	1.2	&	26.8	$\pm$	2.6	&	304	$\pm$	10	\\
PN\, A66\,31	&	-23.2	$\pm$	0.4	&	152.3	$\pm$	2.8	&	-0.8	$\pm$	0.1	&	40.1	$\pm$	0.6	&	30.3	$\pm$	0.8	&	285	$\pm$	9	\\
NGC\,1360	&	-43.7	$\pm$	1.4	&	209.6	$\pm$	14.8	&	-8.4	$\pm$	1.2	&	50.8	$\pm$	1.4	&	38.8	$\pm$	2.5	&	323	$\pm$	8	\\
EGB\,6	&	-29.7	$\pm$	4.1	&	191.5	$\pm$	416.8	&	-33.9	$\pm$	4.1	&	45.6	$\pm$	4.6	&	20.1	$\pm$	4.0	&	551	$\pm$	61	\\
PN\, LoTr\,1	&	-15.6	$\pm$	1.0	&	203.1	$\pm$	20.3	&	7.8	$\pm$	8.3	&	25.1	$\pm$	3.0	&	6.6	$\pm$	0.6	&	886	$\pm$	66	\\
NGC\,2438	&	-42.1	$\pm$	1.9	&	146.7	$\pm$	7.3	&	-1.9	$\pm$	0.5	&	57.0	$\pm$	1.9	&	60.5	$\pm$	1.9	&	53	$\pm$	9	\\
PN\, A66\,15	&	50.4	$\pm$	4.3	&	361.9	$\pm$	78.0	&	55.0	$\pm$	7.5	&	192.0	$\pm$	35.2	&	27.6	$\pm$	1.8	&	1671	$\pm$	350	\\
PN\, A66\,33	&	-91.1	$\pm$	4.4	&	174.4	$\pm$	34.5	&	17.5	$\pm$	4.6	&	93.4	$\pm$	4.4	&	45.5	$\pm$	3.3	&	569	$\pm$	33	\\
PG\, 1034+001	&	-82.1	$\pm$	0.8	&	158.9	$\pm$	4.1	&	18.0	$\pm$	1.7	&	88.0	$\pm$	0.8	&	25.8	$\pm$	1.2	&	143	$\pm$	1	\\
PN\, Lo\,4	&	-66.8	$\pm$	11.5	&	151.3	$\pm$	101.0	&	-40.6	$\pm$	8.1	&	85.1	$\pm$	13.3	&	14.8	$\pm$	16.0	&	505	$\pm$	84	\\
NGC\,2899	&	6.8	$\pm$	0.8	&	193.4	$\pm$	60.3	&	-26.0	$\pm$	1.5	&	28.2	$\pm$	1.6	&	24.1	$\pm$	0.8	&	132	$\pm$	7	\\
PN\,K\,1-27	&	71.7	$\pm$	82.7	&	322.2	$\pm$	49.9	&	10.0	$\pm$	77.5	&	155.2	$\pm$	42.8	&	70.1	$\pm$	61.7	&	2089	$\pm$	305	\\
PN\,A66\,35	&	-18.0	$\pm$	0.8	&	166.5	$\pm$	8.1	&	0.6	$\pm$	0.0	&	25.8	$\pm$	0.9	&	27.1	$\pm$	0.6	&	88	$\pm$	4	\\
PN\,A66\,36	&	61.9	$\pm$	2.8	&	198.3	$\pm$	54.0	&	35.7	$\pm$	3.3	&	72.7	$\pm$	3.0	&	21.5	$\pm$	3.5	&	279	$\pm$	7	\\
Tc\,1	&	-114.5	$\pm$	7.1	&	42.2	$\pm$	6.5	&	-29.6	$\pm$	5.5	&	185.4	$\pm$	17.6	&	82.0	$\pm$	3.8	&	614	$\pm$	84	\\

\hline
\end{tabular}
\end{table*}

\begin{table*}
\centering
\caption{The average kinematical parameters of IPNe located in the Galactic thin and thick-disk} \label{Table4}
\begin{tabular}{lccc}
\hline
	&	Thin-disk	&	Thick-disk	& The entire sample	\\
\hline
\#	&	19 PNe	&	19 PNe	&	46 PNe	\\
Z (pc)	&	263	$\pm$	23	&	600	$\pm$	90	&	395	$\pm$	58	\\
Vs (km/s)	&	35	$\pm$	7	&	111	$\pm$	13	&	70	$\pm$	21	\\
Vp (km/s)	&	32	$\pm$	6	&	46	$\pm$	9	&	37	$\pm$	6	\\
$V_{LSR}$ (km/s)	&	180	$\pm$	75	&	158	$\pm$	34	&	170	$\pm$	50	\\

\hline
\end{tabular}
\end{table*}

\begin{table*}
\centering
\caption{The position angle of CS proper motion of the PN mimic the structure of IPNe.} \label{Table5}
\scalebox{0.60}{
\begin{tabular}{llcccclllllll}
\hline
PN	&	Gaia EDR3  	&	\multicolumn{5}{c}{Galactic and equatorial coordinates}		&	$\pi_{\rm Gaia}$ &	$\mu_{\alpha}$ &	$\mu_{\delta}$ 	&	PA($^{\rm o}$) &	Direction & Reference\\  \cline{3-7}
 	&	designation	& l	& b		&&	$\alpha$	&	$\delta$	&	mas	&	mas/yr	&	mas/yr	& 	&		&	\\
\hline
CN 1-5	&	4046842607233523840	&	2.29	&	-9.48	&	&	277.30	&	-31.50	&	0.26	$\pm$	0.09	&	-2.10	$\pm$	0.10	&	-4.56	$\pm$	0.07	&	205	$\pm$	9	&	NE	&	\citet{Soker97}	\\
DHW 1-2	&	4335587859735067264	&	11.40	&	17.99	&	&	256.73	&	-9.78	&	0.25	$\pm$	0.09	&	-2.26	$\pm$	0.10	&	-5.26	$\pm$	0.08	&	207	$\pm$	9	&	W	&	\citet{Kerber97}	\\
PN M 1-46	&	4103910524954236928	&	16.45	&	-1.98	&	&	276.98	&	-15.55	&	0.40	$\pm$	0.02	&	1.56	$\pm$	0.02	&	-0.66	$\pm$	0.01	&	62	$\pm$	0	&	NW	&	\citet{Corradi03}	\\
PN M 2-40	&	4161093027951979776	&	24.12	&	3.89	&	&	275.35	&	-6.03	&	0.13	$\pm$	0.05	&	-0.24	$\pm$	0.05	&	-4.75	$\pm$	0.04	&	180	$\pm$	40	&	E	&	\citet{Wareing07}	\\
DeHt 2	&	4376331092036268032	&	27.65	&	16.92	&	&	265.42	&	3.12	&	0.56	$\pm$	0.04	&	-4.90	$\pm$	0.04	&	-8.95	$\pm$	0.03	&	216	$\pm$	2	&	NE	&	\citet{Wareing07}	\\
NGC 6772	&	4261038467432411776	&	33.16	&	-6.39	&	&	288.65	&	-2.71	&	1.09	$\pm$	0.18	&	1.68	$\pm$	0.25	&	-0.52	$\pm$	0.24	&	30	$\pm$	3	&	E 	&	\citet{Ramos09}	\\
PN G050.6+00.0	&	4321579226909973376	&	50.67	&	0.01	&	&	291.15	&	15.74	&	0.65	$\pm$	0.12	&	-1.81	$\pm$	0.11	&	-3.61	$\pm$	0.11	&	272	$\pm$	17	&	SE	&	\citet{Sabin08}	\\
PM 1-295	&	4514868732516293760	&	51.36	&	1.81	&	&	289.83	&	17.20	&	0.39	$\pm$	0.02	&	-3.30	$\pm$	0.02	&	-3.07	$\pm$	0.02	&	278	$\pm$	1	&	N	&	\citet{Guerrero98}	\\
NGC 6891	&	1803234906762692736	&	54.20	&	-12.11	&	&	303.79	&	12.70	&	0.37	$\pm$	0.03	&	1.17	$\pm$	0.04	&	-5.33	$\pm$	0.03	&	128	$\pm$	4	&	SW	&	\citet{Wareing07}	\\
NGC 6853	&	1827256624493300096	&	60.84	&	-3.70	&	&	299.90	&	22.72	&	2.57	$\pm$	0.04	&	10.52	$\pm$	0.03	&	3.70	$\pm$	0.04	&	45	$\pm$	0	&	NW	&	\citet{Wareing07}	\\
NGC 7094	&	1770058865674512896	&	66.78	&	-28.20	&	&	324.22	&	12.79	&	0.59	$\pm$	0.03	&	3.68	$\pm$	0.03	&	-10.53	$\pm$	0.03	&	142	$\pm$	1	&	NW	&	\citet{Soker97}	\\
PN A66 61	&	2127684982639844224	&	77.70	&	14.78	&	&	289.79	&	46.25	&	0.62	$\pm$	0.06	&	-0.24	$\pm$	0.08	&	-1.10	$\pm$	0.08	&	18	$\pm$	6	&	W	&	\citet{Tweedy94b}	\\
MWP 1	&	1855295171732158080	&	80.36	&	-10.41	&	&	319.28	&	34.21	&	1.97	$\pm$	0.04	&	-5.49	$\pm$	0.03	&	10.43	$\pm$	0.03	&	345	$\pm$	2	&	S	&	\citet{Rauch99}	\\
PN A66 78	&	1850685091269441792	&	81.30	&	-14.91	&	&	323.87	&	31.70	&	0.59	$\pm$	0.03	&	-3.02	$\pm$	0.02	&	0.43	$\pm$	0.02	&	357	$\pm$	2	&	SE	&	\citet{Soker97}	\\
Jacoby 1	&	1595941441250636672	&	85.37	&	52.35	&	&	230.44	&	52.37	&	1.30	$\pm$	0.04	&	-3.79	$\pm$	0.05	&	11.35	$\pm$	0.05	&	354	$\pm$	4	&	S	&	\citet{Tweedy96}	\\
SH 1-89 	&	2164519416761365248	&	89.84	&	-0.64	&	&	318.53	&	47.77	&	0.53	$\pm$	0.40	&	-1.32	$\pm$	0.52	&	-2.71	$\pm$	0.45	&	56	$\pm$	22	&	SW	&	\citet{Hsia20}	\\
PN DeHt 5	&	2229624931896924160	&	111.09	&	11.64	&	&	334.89	&	70.93	&	2.97	$\pm$	0.04	&	-13.70	$\pm$	0.04	&	-18.83	$\pm$	0.05	&	204	$\pm$	1	&	NW	&	\citet{Borkowski90}	\\
PN BV 1	&	430620640942643968	&	119.37	&	0.33	&	&	5.00	&	62.98	&	0.39	$\pm$	0.20	&	-2.49	$\pm$	0.18	&	-0.82	$\pm$	0.20	&	87	$\pm$	11	&	SE	&	\citet{Soker97}	\\
PN A66 6 	&	467936205865972352	&	136.10	&	4.93	&	&	44.67	&	64.50	&	0.85	$\pm$	0.14	&	3.16	$\pm$	0.06	&	-5.98	$\pm$	0.13	&	162	$\pm$	4	&	SW	&	\citet{Soker97}	\\
PN M 2-2  	&	469056092819323392	&	147.87	&	4.20	&	&	63.31	&	56.95	&	0.19	$\pm$	0.05	&	0.29	$\pm$	0.06	&	1.87	$\pm$	0.05	&	32	$\pm$	7	&	S	&	\citet{Wareing07}	\\
PN IsWe 1	&	250358801943821952	&	149.71	&	-3.40	&	&	57.27	&	50.00	&	2.33	$\pm$	0.06	&	18.15	$\pm$	0.05	&	-7.64	$\pm$	0.05	&	97	$\pm$	0	&	N-NE	&	\citet{Xilouris96}	\\
SH 2-216	&	254092090595748096	&	158.49	&	0.47	&	&	70.84	&	46.70	&	7.84	$\pm$	0.05	&	21.62	$\pm$	0.05	&	-13.02	$\pm$	0.04	&	28	$\pm$	0	&	E	&	\citet{Borkowski90}	\\
PN WeDe 1 	&	3341996555048041856	&	197.41	&	-6.44	&	&	89.85	&	10.69	&	1.71	$\pm$	0.10	&	-0.04	$\pm$	0.12	&	-1.38	$\pm$	0.09	&	16	$\pm$	50	&	NW	&	\citet{Borkowski90}	\\
PN\, K\,2-2	&	3158419684195782656	&	204.15	&	4.73	&	&	103.10	&	9.97	&	1.19	$\pm$	0.05	&	-2.26	$\pm$	0.05	&	-8.39	$\pm$	0.04	&	190	$\pm$	4	&	SW	&	\citet{Wareing07}	\\
PN A66 21	&	3163546505053645056	&	205.14	&	14.24	&	&	112.26	&	13.25	&	1.67	$\pm$	0.07	&	-2.93	$\pm$	0.07	&	-8.48	$\pm$	0.06	&	184	$\pm$	5	&	SE	&	\citet{Tweedy96}	\\
HDW 5	&	3001563840710096512	&	218.99	&	-10.78	&	&	95.90	&	-10.22	&	0.91	$\pm$	0.04	&	-0.25	$\pm$	0.05	&	-13.55	$\pm$	0.05	&	171	$\pm$	31	&	N	&	\citet{Ali99}	\\
PFP 1	&	3058094200264637312	&	222.13	&	3.91	&	&	110.57	&	-6.36	&	1.85	$\pm$	0.07	&	-3.77	$\pm$	0.07	&	-8.61	$\pm$	0.06	&	186	$\pm$	3	&	NW	&	\citet{Pierce04}	\\
We 1-6	&	3047204259149524480	&	224.94	&	1.06	&	&	109.36	&	-10.18	&	0.69	$\pm$	0.04	&	2.04	$\pm$	0.04	&	-2.32	$\pm$	0.03	&	132	$\pm$	2	&	N	&	\citet{Soker97}	\\
NGC 3242	&	5668874905325843456	&	261.05	&	32.05	&	&	156.19	&	-18.64	&	0.73	$\pm$	0.05	&	-4.60	$\pm$	0.05	&	0.83	$\pm$	0.05	&	250	$\pm$	2	&	NE	&	\citet{Ramos09}	\\
PN K 1-28	&	5455506530699796608	&	270.17	&	24.84	&	&	158.63	&	-29.19	&	0.47	$\pm$	0.07	&	-6.98	$\pm$	0.06	&	-3.74	$\pm$	0.07	&	188	$\pm$	2	&	N	&	\citet{Soker97}	\\
NGC 2867	&	5300450617836957312	&	278.16	&	-5.94	&	&	140.36	&	-58.31	&	0.33	$\pm$	0.06	&	-5.50	$\pm$	0.07	&	3.04	$\pm$	0.06	&	115	$\pm$	2	&	S	&	\citet{Soker97}	\\
PN\,MeWe\,1-2	&	5258596741332679040	&	283.45	&	-1.39	&	&	153.60	&	-58.20	&	0.43	$\pm$	0.19	&	-8.17	$\pm$	0.21	&	2.69	$\pm$	0.18	&	97	$\pm$	3	&	SW	&	\citet{Kerber2000}	\\
ESO 320-28 	&	5380413425574962944	&	291.43	&	19.26	&	&	178.12	&	-42.29	&	0.13	$\pm$	0.07	&	-6.79	$\pm$	0.05	&	-2.25	$\pm$	0.04	&	189	$\pm$	1.00	&	E	&	\citet{Soker97}	\\

\hline
\end{tabular}}
\end{table*}

\bibliographystyle{pasa-mnras}
\bibliography{PN-ISM-III}
\end{document}